\documentclass{emulateapj}

\def\sun{\ifmmode\odot\else$\odot$\fi}
\def\HII{\hbox{H\,{\sc ii}}}

\def\H2{\hbox{H$_{2}$}}

\def\Halpha{\hbox{\rm H$\alpha$}}

\def\Paalpha{\hbox{\rm Pa$\alpha$}}

\def\Msun{M$_{\odot}$}
\def\Lsun{L$_{\odot}$}

\def\kms{${\rm km~s}^{-1}$}

\shorttitle{Understanding the 8$\micron$ vs. Pa$\alpha$ relationship}
\shortauthors{D\'{\i}az-Santos et al.}

\begin{document}

\title{Understanding the 8$\micron$ $\lowercase{\rm vs.}$ P$\lowercase{\rm a} \alpha$ relationship on sub-arcsecond scales in Luminous Infrared Galaxies$\,^{\lowercase{\rm a,b}}$}


\author{Tanio~D\'{\i}az-Santos\altaffilmark{1},  
Almudena~Alonso-Herrero\altaffilmark{1}, Luis~Colina\altaffilmark{1},
        Christopher~Packham\altaffilmark{2},
        James~T.~Radomski\altaffilmark{3},
	and Charles~M.~Telesco\altaffilmark{2}}
\altaffiltext{1}{Based on observations obtained with T-ReCS instrument at the
        Gemini South Observatory, which is operated by AURA, Inc., under a
        cooperative agreement with the NSF on behalf of the Gemini
        partnership: NSF (United States), PPARC (UK), NRC (Canada), CONICYT
        (Chile) ARC (Australia), CNPq (Brazil) and CONICET (Argentina).}
\altaffiltext{b}{Also
        based on observations with the NASA/ESA Hubble Space Telescope,
        obtained from the data archive at the Space Telescope Science
        Institute, which is operated by the association of universities for
        research in astronomy, inc., under NASA contract NAS~5-266555.} 
\altaffiltext{1}{Departamento de Astrof\'{\i}sica Molecular e Infrarroja,
        Instituto de Estructura de la Materia (IEM), CSIC, Serrano 121, 
E-28006 Madrid, Spain}
\altaffiltext{2}{Department of Astronomy, University of Florida, 211 Bryant Science Center, P.O. Box 112055, Gainesville, FL 32611-2055}
\altaffiltext{3}{Gemini Observatory, c/o AURA, Casilla 603, La Serena, Chile}

\begin{abstract}
This work explores in detail the relation between the 8$\,\micron$ and
the \Paalpha\ emissions for 122 \HII\ regions identified in a sample of 10 
low-$z$ LIRGs with nearly constant metallicity
($12 + \log ({\rm O/H}) \sim 8.8$).
We use Gemini/T-ReCS high-spatial resolution
($\lesssim 0\farcs4\sim 120\,$pc
for the average distance of 60\,Mpc of our sample)
mid-infrared imaging (at $8.7\,\mu$m or $10.3\,\mu$m) 
together with {\it HST}/NICMOS continuum and 
Pa$\alpha$ images.  The LIRG \HII\ regions extend the 
$L_{\rm 8\,\mu m}$ vs. $L_{\rm Pa\alpha}$ relation found for \HII\ knots
in the high-metallicity SINGS galaxies by about two orders of magnitude to
higher luminosities.
Since the metallicity of the LIRG sample is nearly constant,
we can rule out this effect as a cause for the scatter seen
in the relationship. In turn, it is attributed to two effects:
age and PAH features.
The $L_{\rm 8\,\mu m}$/$L_{\rm Pa\alpha}$ ratio, which varies by
a factor of ten
for the LIRG \HII\ regions, is reproduced by a model with instantaneous 
star formation and ages ranging from $\sim\,$4 to 7.5\,Myr.
The remaining dispersion around the model predictions  
for a given age is probably due to differential contributions of the PAH features 
 (the $8.6\,\mu$m, in
our case) to the $8\,\mu$m emission from galaxy to galaxy.

\end{abstract}


\section[]{Introduction}\label{s:intro}

The $8\,\mu$m luminosity is potentially one of the most interesting
star formation rate (SFR) indicators as it can be used for sources
identified  in deep infrared (IR) {\it Spitzer}/MIPS surveys at 
24$\,\micron$ which at z\,$\sim$\,2 corresponds 
to rest-frame 8$\,\micron$ emission. 
The MIPS $24\,\mu$m (or IRAS $25\,\mu$m) luminosities
appear to be well correlated with the number of ionizing photons
as derived from extinction corrected Pa$\alpha$ and H$\alpha$ luminosities
at least at high metallicities 
(\citealt{Wu05}; \citealt{Cal05}; \citealt{Calzetti07}, Cal07 hereafter;
\citealt{AAH06a}, AAH06a  hereafter). 
The situation  for the 
$8\,\mu$m emission is less clear (Cal07, 
\citealt{AAH06b}, AAH06b hereafter). In particular, AAH06b
found that the individual \HII\ regions and the integrated emission
of Luminous Infrared Galaxies (LIRGs,
$L_{{\rm IR}}=10^{11}-10^{12}\,{\rm L}_\odot$)
show a different behavior in the $L_{\rm 8\,\mu m}$ vs.
$L_{\rm Pa\alpha}$ relation, and suggested that only the integrated properties
of galaxies should be used when calibrating the SFR in 
terms of the $8\,\mu$m luminosity (see also \citealt{Wu05}). Recently 
Cal07 for \HII\ knots identified in the SINGS 
(\textit{Spitzer} IR Nearby
Galaxies Survey, \citealt{Kennicutt03}) galaxies
concluded that the larger scatter of the $8\,\mu$m vs. Pa$\alpha$ 
relation is due to the combined effects of extinction, 
metallicity and the star formation history of the regions.

The emission in the 8$\,\micron$ spectral region is produced by 
thermal continuum from hot dust as well as by Polycyclic Aromatic
Hydrocarbon (PAH) feature emission. PAH are 
commonly observed in the MIR spectra
of local (e.g., Roche et al. 1991; \citealt{Lutz98}; \citealt{Genzel98};
\citealt{Brandl06}; \citealt{Smith07}) and high-$z$
star-forming galaxies (\citealt{Sajina07}). However, while the 
dust continuum emission as traced by the MIPS $24\,\mu$m emission 
is found to be more peaked in \HII\ regions, 
the $8\,\mu$m (mostly as PAH) emission arises from
Photo-dissociation Regions (PDR) 
(\citealt{Helou04};  \citealt{Bendo06}; \citealt{Povich07}; 
\citealt{Lebouteiller07}). This implies that the 
PAH carriers can also be excited by the galaxy field
radiation (\citealt{Peeters04}; \citealt{TG05}) 
not directly associated with young ionizing stellar populations,
explaining why the 8$\,\micron$ emission appears to be 
more extended and diffuse than the \Halpha\ or
\Paalpha\ emission (\citealt{Helou04}: \citealt{Cal05}; AAH06b; 
\citealt{Engelbracht06}).
Consequently, aperture effects may have important implications
when measuring the emission from individual star-forming regions.

In this work we further study the 8$\,\micron$ emission 
at sub-arcsecond scales using observations obtained with the 
Thermal-Region Camera Spectrograph (T-ReCS; \citealt{Telesco98}) on Gemini
South of  a sample of low-$z$ LIRGs (see also AAH06a, AAH06b).
In particular, we explore the effects of the age and extinction 
of the individual star-forming regions  on 
the 8$\,\micron$ vs. \Paalpha\ relation and how they may contribute
to the observed scatter of the relation.
We also compare our results with those of 
Cal07 for high-metallicity \HII\ knots in star-forming
galaxies drawn from the SINGS sample.
The paper is organized as follows: In \S2, the sample, observations,
and data reduction are presented. \S3 describes the analysis of
the data. The overall morphology of the LIRGs is presented in \S4.
\S5 analyzes the
$L_{\rm 8\,\mu m}$/$L_{\rm Pa\alpha}$ relationship in detail.
The summary of the results is given in \S6. We use 
$H_0=70\,{\rm km\,s}^{-1}{\rm Mpc}^{-1}$, $\Omega_{\rm M}=0.27$, and
$\Omega_\Lambda=0.73$.


\section[]{Observations and Data Reduction}\label{s:obs}

\subsection[]{The Sample}\label{ss:sample}

We have obtained MIR imaging of a  total of ten LIRGs (see Table~1 for
details)  taken from 
the complete, volume-limited sample 
of local LIRGs defined by AAH06a. 
The full sample of AAH06a was   drawn from the $IRAS$ Revised 
Bright Galaxy Sample (\citealt{Sanders03}) and 
selected such that the \Paalpha\ emission
line ($\lambda_{\rm rest}$\,=\,1.875\,$\micron$) could be observed with
the NICMOS F190N filter on the
\textit{Hubble Space Telescope} (\textit{HST}). This restriction 
means that the full sample
 is limited to nearby galaxies ($d<75\,$Mpc). The ten galaxies  
studied in this paper represent the
majority of the LIRGs in the sample that can be observed from the southern
hemisphere. 
Four galaxies were previously studied  by AAH06b but are
fully re-analyzed and included here for completeness and consistency.
The metallicities of the sample, taken from 
 \citealt{Relano07} and \citealt{Vacca92},
are on average $12 + \log  {\rm (O/H)} =8.8$ (see Table~1 for individual
values).


\subsection[]{MIR Imaging Observations}\label{ss:obs}

The MIR observations of the LIRGs were obtained with T-ReCS 
(\citealt{Telesco98}) on the Gemini South telescope
during semesters 2005B, 2006A and 2006B (program IDs: GS-2005B-Q-10,
GS-2006A-Q-7 and GS-2006B-Q-9, respectively).
The data from the first semester were taken with the
broad-band $N$ filter ($\lambda_{\rm c}$\,=\,10.36\,$\micron$;
$\Delta\lambda$\,=\,5.27\,$\micron$), whereas data obtained in 
2006 used the narrow-band Si-2 filter
($\lambda_{\rm c}$\,=\,8.74\,$\micron$; $\Delta\lambda$\,=\,0.78\,$\micron$).
Detailed information about the observations is given in Table~\ref{t:obs}.
The 320\,$\times$\,240 pixel detector
together with its 0\farcs09\,pixel$^{-1}$ plate scale yield a field of
view (FOV) of $\sim$\,28.5\arcsec\,$\times$\,21.5\arcsec. At the central
wavelengths of the $N$ and Si-2 filter band-passes, the telescope/instrument
system provides resolutions of $\sim$\,0\farcs32 and 0\farcs27,
respectively. The observing conditions (seeing FWHM of 
$0\farcs3- 0\farcs4$, 
Table~\ref{t:obs}) indicate that the observations were almost diffraction
limited.

The observations were done in a standard chop-nod mode
to remove the time-variable sky background, telescope thermal emission,
and the 1/f detector noise. The chop throw was 15$\,^\circ$ and the 
telescope
nodding was performed every 30\,s in all cases (see also \citealt{Packham05}).
The observations were divided in two datasets to avoid 
observing problems or a sudden change in weather conditions.
We observed a Cohen standard star (appropriate for
MIR observations, \citealt{Cohen99}) for each galaxy to obtain
the absolute flux calibration of the images. The standard star observations
were always taken with the same instrument configuration and immediately 
before or after the target to minimize the difference in airmass 
and hence to reduce photometric uncertainties.
The large-scale morphology of some of the targets made it 
necessary to rotate the detector along the major axis 
of the galaxy. In those cases the standard stars
were observed with the same configuration.

\begin{figure*}
\epsscale{1.1}
\plotone{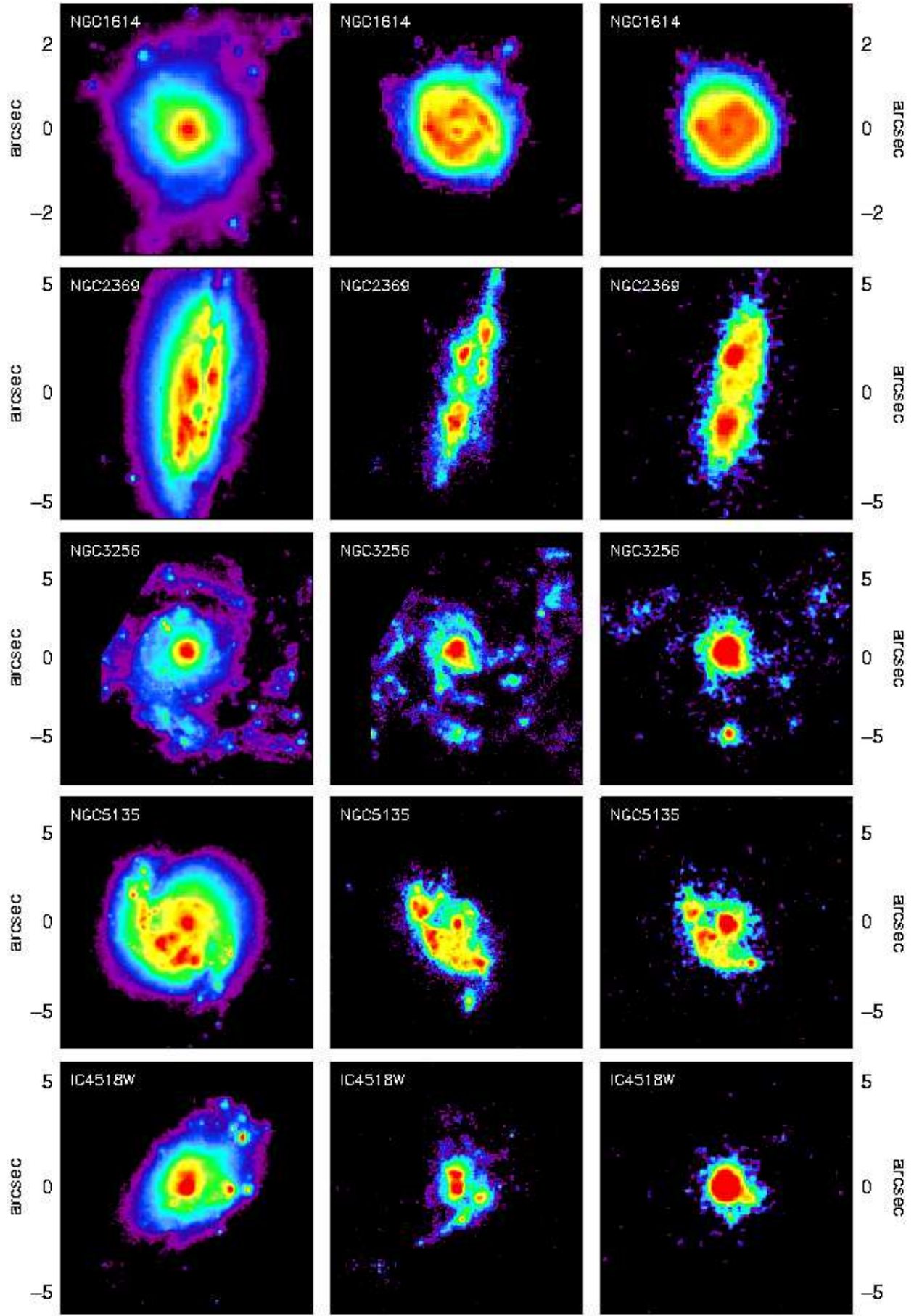}
\vspace{.25cm}
\caption{\footnotesize {\it HST}/NICMOS 1.6\,$\micron$ continuum
  (left), and 
continuum-subtracted \Paalpha\ line (center) full spatial resolution images
(see also AAH06a), and 
T-ReCS MIR (8.7\,$\micron$ or $N$-band) images (right). All the images are 
displayed on a logarithmic scale.
 North is up, East to the left. The displayed FOV is optimized to show the
  extent of the MIR emission. 
 [\textit{See the electronic edition of the Journal for a color version of this figure.}]}\label{f:panel1}
\end{figure*}

\setcounter{figure}{0}
\begin{figure*}
\epsscale{1.1}
\plotone{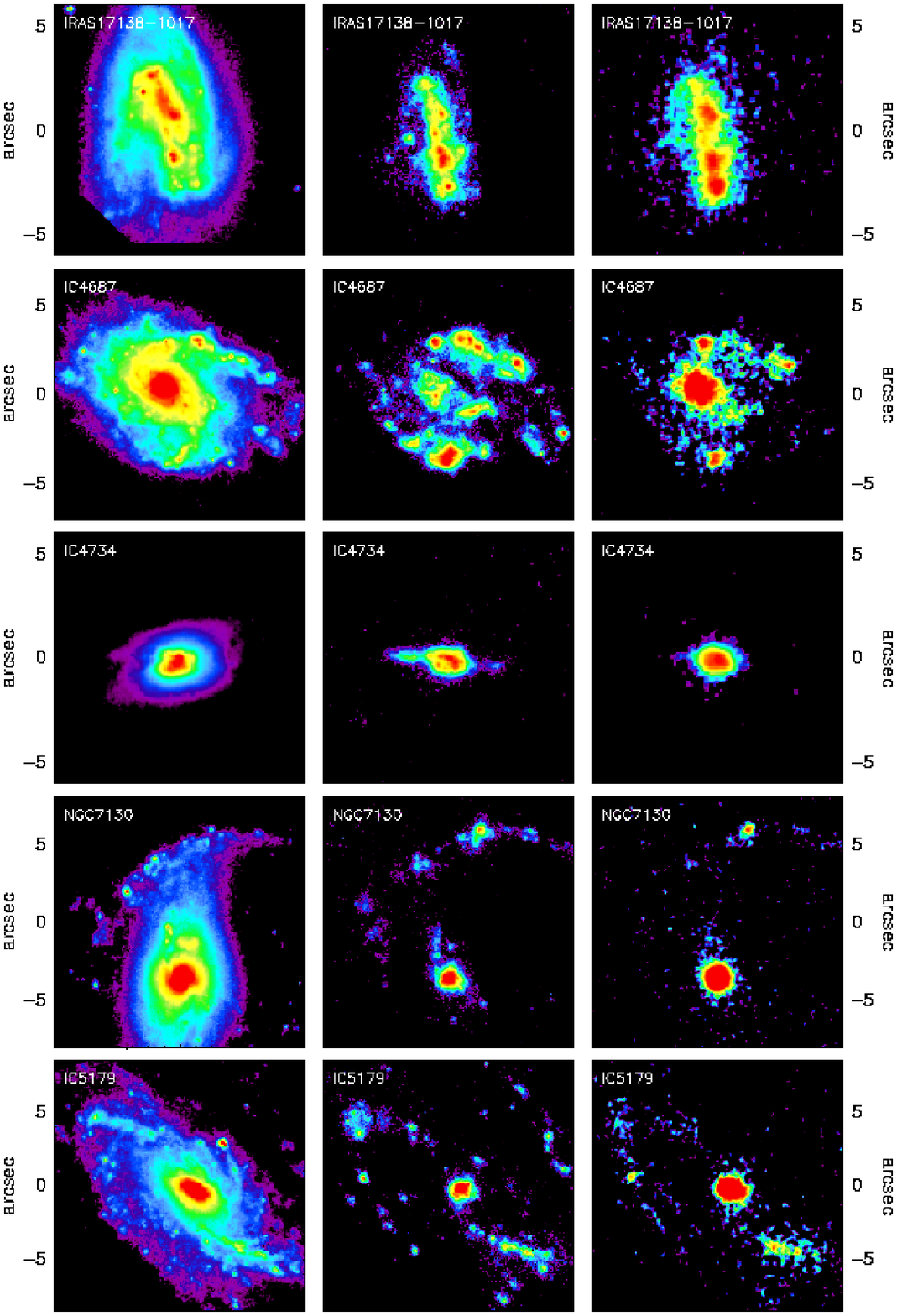}
\vspace{.25cm}
\caption{\footnotesize Continued. [\textit{See the electronic edition of the Journal for a color version of this figure.}]}
\end{figure*}

\begin{figure*}
\epsscale{1.2}
\plotone{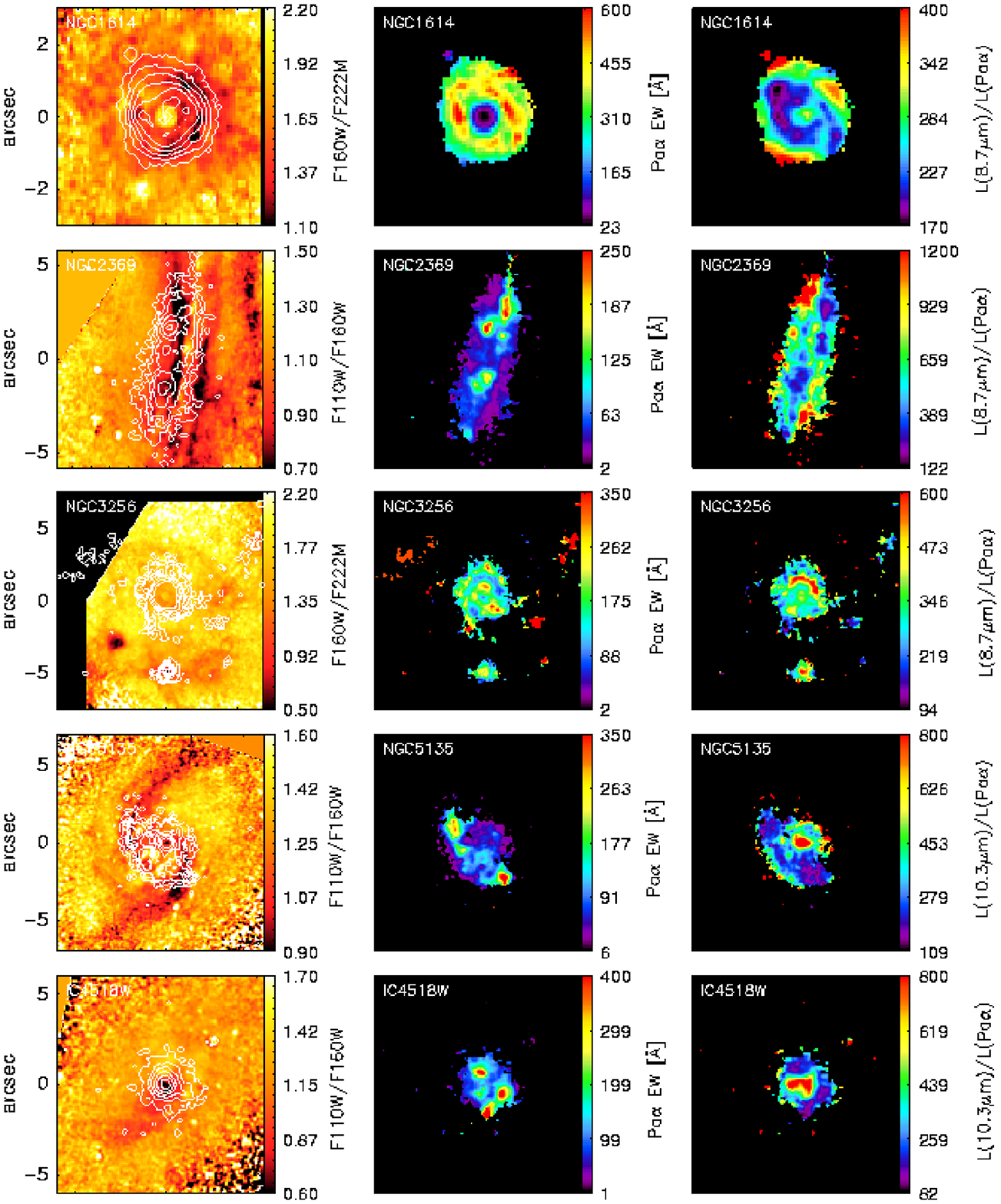}
\vspace{.25cm}
\caption{\footnotesize The left and the middle panels are the {\it HST}/NICMOS
  F110W/F160W ratio 
(or F160W/F222M
ratio for NGC~1614 and NGC~3256) and  \Paalpha\ EW images, respectively, smoothed to the T-ReCS 
resolution (see \S2.5). The contours superimposed on the left panel are the T-ReCS/MIR emission (see Fig.~\ref{f:panel1}). The right panel are the observed 
maps of the MIR/\Paalpha\ ratios. [\textit{See the electronic edition of the Journal for a color version of this figure.}]}\label{f:panel2}
\end{figure*}

\setcounter{figure}{1}
\begin{figure*}
\epsscale{1.2}
\plotone{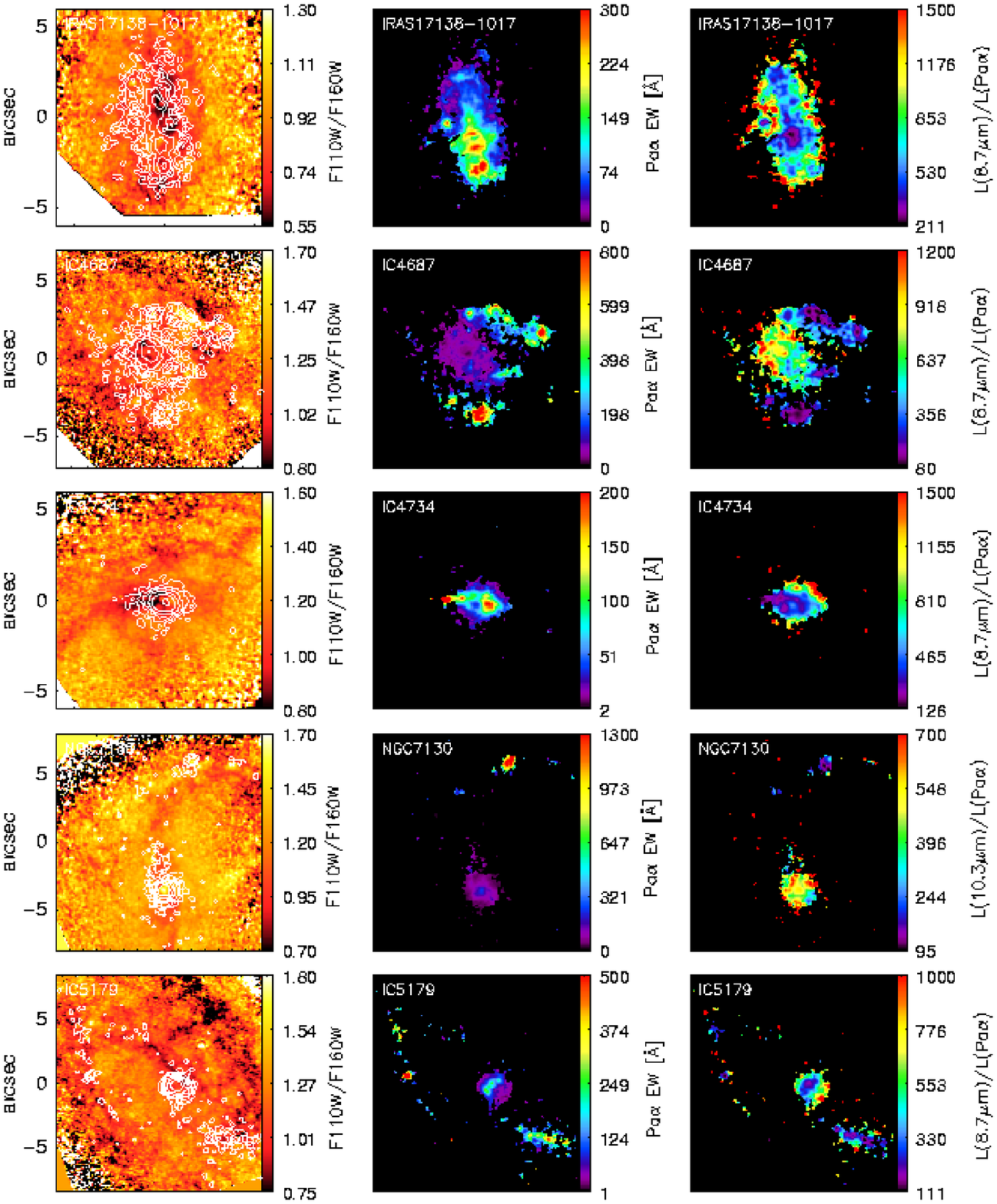}
\vspace{.25cm}
\caption{\footnotesize Continued. [\textit{See the electronic edition of the Journal for a color version of this figure.}]}
\end{figure*}

\subsection[]{Data Reduction Procedure}\label{ss:reduc}

The data are stored automatically in \textit{savesets} (indivisible subsets
of data) that can be accessed for visual inspection for each chop/nod
position. This allows to discard images
 affected by any type of instrumental
noise pattern (e.g., narrow diagonal stripes
of increased signal across the detector or over-variations of the
background flux along the array or in between chop sets). There
were no bad 
\textit{savesets} in our case.
Next, the backgrounds were removed and the galaxy  images coadded to obtain a single image of the target.
This was done for both datasets in which the observation was divided
(see above). These and the following procedures
were also applied to the standard star.

We applied a flat-field image
for each dataset which was constructed using the average of all the sky
pointings taken at each chop integration. The target datasets were then
averaged in a single image from which the
residual background flux was subtracted.
This step requires a special treatment because the final images with
long integration times can suffer from a residual noise pattern
that introduces low-level flux variations along the short axis
of the detector (broad wave-like horizontal stripes). To 
subtract the residual background, the image was first fitted
to a 2D plane with a fitting procedure that iteratively rejects
outliers 2.25\,$\sigma$ above/below this plane. At the same time,
preselected locations with galaxy emission were masked out.
Pixels within these areas with 1.5\,$\sigma$ fluxes above/below 
the 2D fit were not used in following steps. Next, each row of this
2D-subtracted, masked image was smoothed and fitted to a 1D spline.
By combining all these 1D fits, a residual background 2D-image
was created and subtracted from the object image. All these steps
improve the quality of the final images and introduce no more than
$\sim$\,10\% uncertainty in the fluxes of the measured regions.
The  T-ReCS images were rotated to the north-east direction and convolved 
 with a Gaussian kernel of $\sigma$\,=\,1\,pixel 
to accurately center the apertures on the faintest
\HII\ regions (see \S3.1).

Fig.~\ref{f:panel1} shows fully reduced MIR images (right) of the sample 
together with the {\it HST}/NICMOS 1.6\,$\micron$ continuum   and 
\Paalpha\, images.

\subsection[]{Photometric calibration}\label{ss:photcal}

The photometric calibration was obtained with the Cohen
standard stars (\citealt{Cohen99}).
Since the Cohen stars  have synthetic flux-calibrated 
spectra, by convolving the telescope throughput and filter bandpass 
with the spectra, one can obtain the flux of the standard stars
in physical units. In addition, aperture photometry was performed
on the standard stars to obtain the count rates, and the flux
conversion factor was calculated as the ratio of both quantities. 
The photometric calibration is accurate to $\lesssim$\,10--15\%. 

The need for a color correction for the broad $N$ filter flux
densities was explored. This factor would account for the difference
in the spectral slope of the standard stars and those of our galaxies.
While the MIR spectrum of a star is similar to that of the
Rayleigh-Jeans part of a black body, the spectrum of an \HII\,
region or a nucleus may differ significantly.
In order to quantify this difference, we used the $Spitzer$/IRS
spectra of several LIRGs of the sample classified as \HII-like
(NGC~1614, NGC~3256 and IRAS~17138-1017). We normalized their
integrated N-band flux densities to those of the T-ReCS reference
standard stars observed for flux calibrating each galaxy. Next,
the spectra were convolved with the same T-ReCS configuration used
for the observations, and the N-band flux densities of the spectra
were calculated. Then, they were compared with those of the standards.
The differences were within 5\%. Given the uncertainties of the
method, we decided not to correct the
flux density of the \HII\ regions for this color difference.
The same approach was employed for calculating the correction
factor of an AGN spectrum. We used a power-law function of
$F_\nu\,\propto\,\nu^{-\alpha}$, with $\alpha\,\approx\,1.75$,
to represent its MIR emission (\citealt{Weedman05}), and obtained
a factor of $\simeq\,$1.15.
The MIR flux densities of the two Sy2 (NGC~5135, IC~4518W) and
one LINER/Sy (NGC~7130) nuclei in the sample observed with the
N-band filter were corrected by this factor.
This kind of color correction is not necessary for the Si-2 filter
as it is relatively narrow.

Since the galaxies were observed with two different MIR filters, 
we investigated whether a conversion factor between
the  $N$-band and the Si-2 filter flux densities 
was needed. To do so, we used the $Spitzer$/IRS low-resolution spectra
of several \HII-classified LIRGs of our sample (NGC~1614, NGC~3256,
IRAS~17138-1017, and IC~4734) integrated over their central
(10\arcsec\,$\times$\,3.7\arcsec) regions. We verified that, within
a $\sim\,$10\% uncertainty, the 8.7$\,\micron$ and $N$-band flux
densities were equivalent. The same was assumed for our individual
\HII\ regions. For the Sy nuclei
we used a conversion factor
$f_\nu\,$[8.7\,\micron]/$f_\nu\,$[$N$-band] of 0.56 
as done by AAH06b.

\subsection[]{HST/NICMOS images}

The sample of LIRG galaxies were observed with the NIC2 camera (pixel size of
 0.075\arcsec) of NICMOS using
two broad-band filters (F110W and F160W) and two narrow-band filters (F187N
and F190N), except for two galaxies (NGC~1614 and NGC~3256) from 
the NICMOS GTO programs
which were observed with  the F160W and F222M filters
(see Alonso-Herrero et al. 2001, 2002). 
The full description of the data reduction of the {\it HST}/NICMOS 
images can be found in AAH06. To compare the NICMOS images with 
the T-ReCS images we performed the following additional steps.
The NICMOS images were: (1) rotated to the 
north-east orientation, (2) re-scaled
to the pixel scale of the T-ReCS detector, (3) shifted to
the same position as the MIR images, (4) smoothed with a Gaussian to
match the MIR resolution (see Table~\ref{t:obs}), and
(5) background subtracted. In addition, we constructed J/H (or H/K)
ratio maps by dividing their F110W and F160W (or F160W and F222M) continuum 
images;
\Paalpha\ equivalent width (EW) maps by dividing the  continuum-subtracted 
\Paalpha\ images
by their adjacent F187N continuum image; and MIR/\Paalpha\ ratio maps by
dividing the T-ReCS images by the continuum-subtracted \Paalpha\
images.
These maps are presented in Fig.~\ref{f:panel2} for our
sample of galaxies.

\begin{figure}
\epsscale{1.1}
\plotone{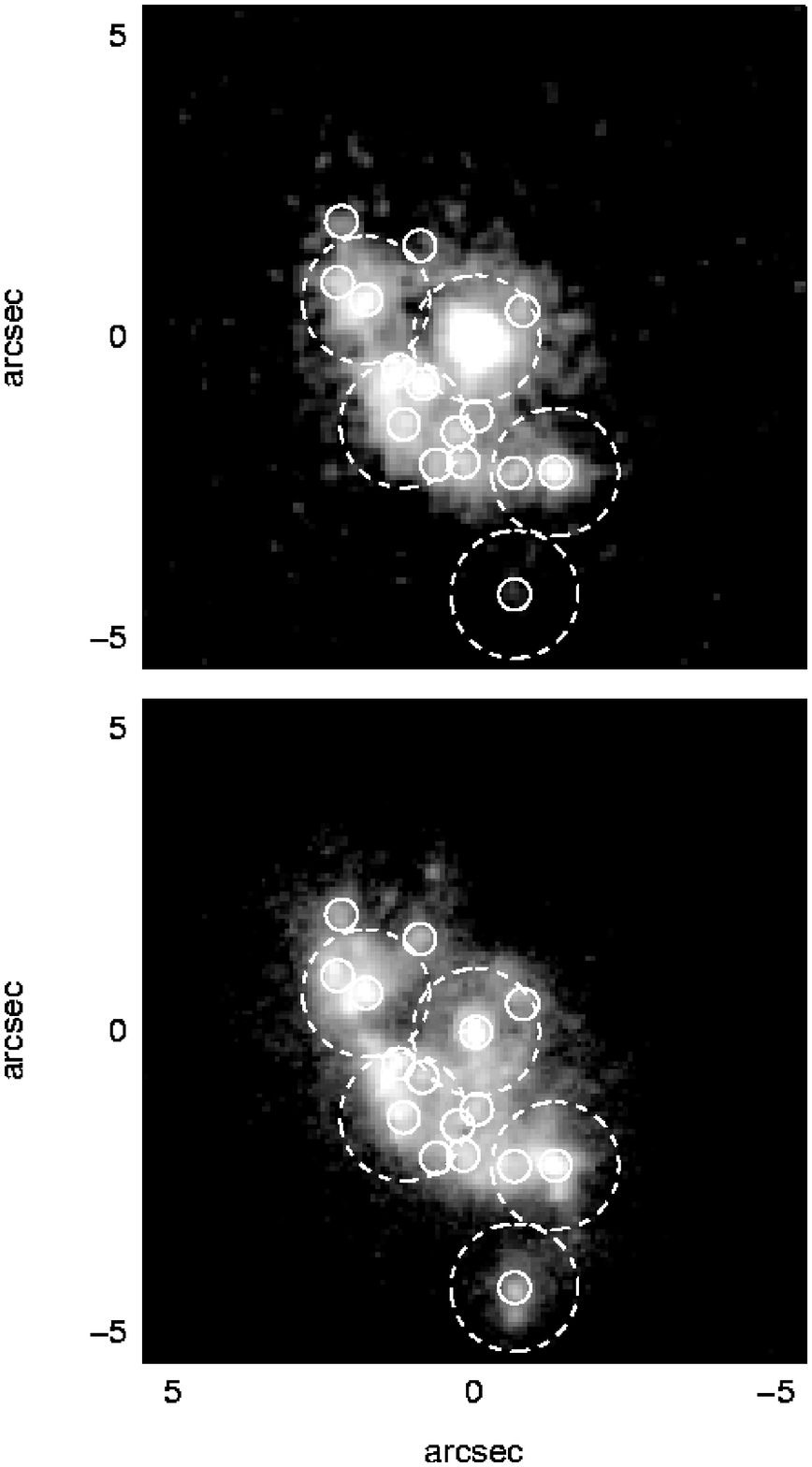}
\caption{\footnotesize T-ReCS $N$-band image (upper panel) and 
{\it HST}/NICMOS continuum-subtracted Pa$\alpha$ image (lower panel) 
of the central region
  ($\sim 10\arcsec \times 10\arcsec$) of NGC~5135. The \HII\ regions selected
  for this LIRG are marked 
as solid circles for the small $r=75\,$pc aperture and as dashed circles for
  the large $r=300\,$pc aperture.}\label{f:apert} 
\vspace{.25cm}
\end{figure}

\section[]{Analysis and Results}\label{s:analysis}

\subsection[]{Selection of the regions and aperture photometry}\label{ss:selec}

We identified the nucleus of each galaxy as the source
with the highest MIR flux density. Although this is straightforward in most of
the galaxies, there are two examples: NGC~2369 and IRAS~17138$-$1017, where
the position of the nucleus is not clear from the NIR continuum images.
For the selection of the \HII\ regions, first we visually identified
them in the T-ReCS images. A second inspection was done 
in the \Paalpha\ images to include sources with high hydrogen recombination
line fluxes but low MIR emission. 
We finally rejected in both
sets sources with integrated flux densities (see below) 
below 2 times the standard
deviation of the background (i.e., a 2$\sigma$ threshold).

Except for some of the most luminous nuclei (NGC~5135, IC~4518W),
the majority of the MIR sources appear extended.
Therefore we used circular aperture photometry
(instead of point source function fitting) to measure the flux
densities of nuclei and star-forming regions of the LIRGs.
We note that AAH06a measured the \Paalpha\ emitting regions with variable
apertures depending on the size of the \HII\ regions instead of with
fixed apertures as done in this work.
We explored the possibility of a contamination for underlying emission
in the MIR images. However, the subtraction of a local background
measurement from the flux densities of the \HII\ and nuclear
regions resulted in negligible changes of the results of the paper.
Therefore we decided to work with the values obtained without
performing any local background subtraction because it might
introduce additional uncertainties.

We chose to use three physical 
apertures for the photometry with radii of 75, 150 and 300\,pc. 
The 75\,pc radius aperture was used to take advantage of the 
high spatial resolution data and to obtain information about
the \HII\ regions on the smallest scales.
At the average distance of our sample  ($\sim$\,60\,Mpc),
the diameter of the smallest aperture corresponds to 
$\sim\,$0\farcs52 ($\sim\,$5.8\,pixels),
which is approximately 1.5 times the seeing of our observations
(see Table~\ref{t:obs}).
Thus, the centering errors (less than a half of a pixel) of the
smallest aperture introduce $<$\,10\% uncertainty
in the measured flux densities of the regions (depending on
their brightness).

The largest aperture was chosen  to compare our results
with those of Cal07 for the high-metallicity \HII\ knots of the 
SIGNS galaxies. They used a fixed angular aperture of 13\arcsec\
in diameter. Almost half of the \HII\ knots in the high-metallicity SINGS
sample are in galaxies located at distances between 8 and 10.5\,Mpc. So
using an average (weighed with the number of 
\HII\ knots) SINGS 
distance of $\sim$\,9.2\,Mpc, their fixed 13\arcsec-diameter apertures
correspond to approximately 580\,pc. 
Our 150\,pc-radius aperture is approximately
intermediate between the small and the large apertures.
In Fig.~\ref{f:apert}
we show an example of the physical areas covered by the smallest and
largest apertures in one LIRG in our sample.

To avoid overlapping among the selected regions when performing aperture
photometry, we sorted
them in flux density and rejected those regions within 0.75 times
the diameter of another selected region(s) with higher flux 
density. 
In addition to the 11 (AGN or \HII) nuclei (two in NGC~3256), 
the final number of  \HII\ regions selected for photometry were 
122, 84, and 49  for the apertures of
$r=75,\, 150$, and 300\,pc, respectively.

\begin{figure}
\epsscale{1.15}
\plotone{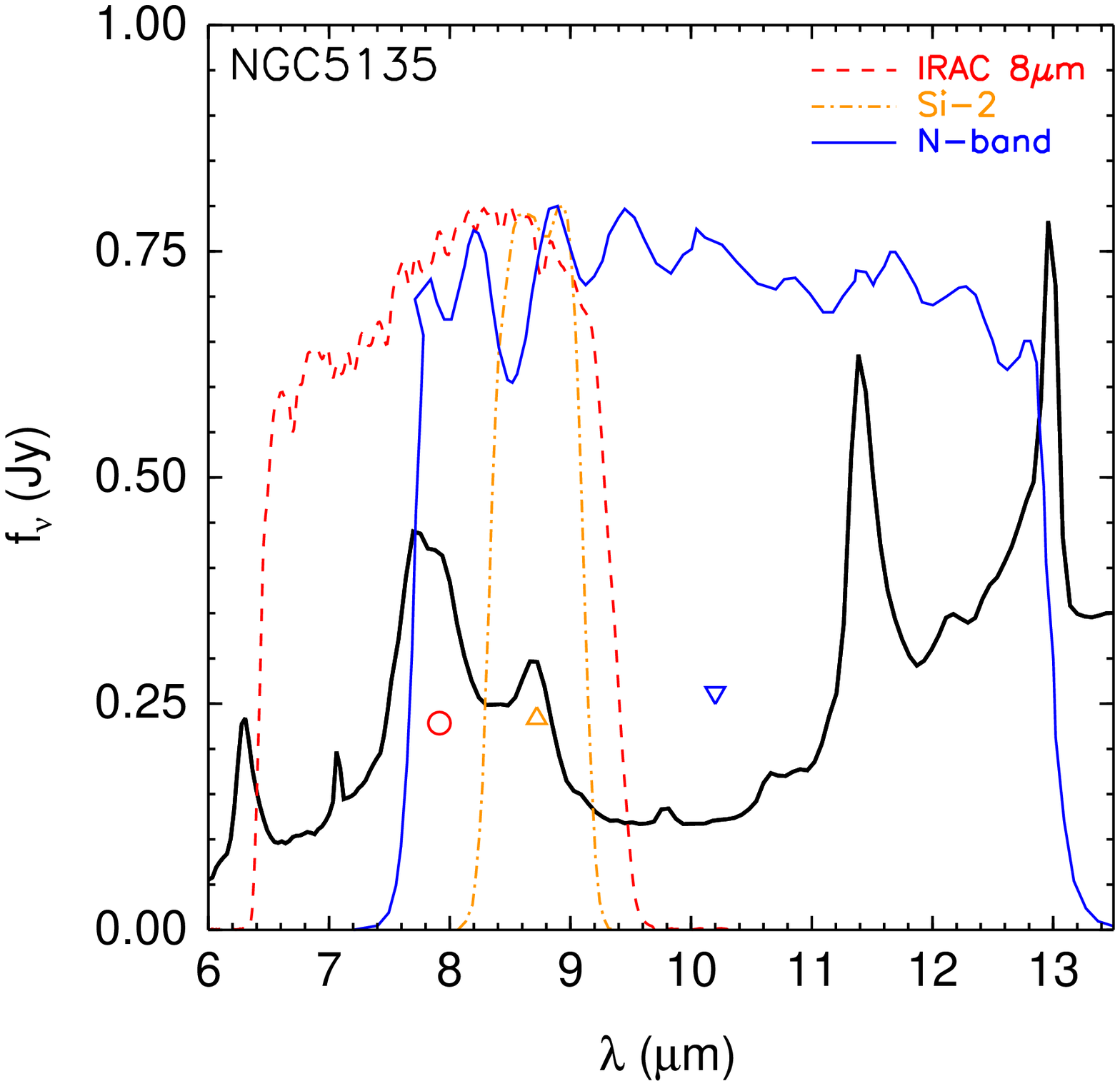}
\caption{\footnotesize $Spitzer$/IRS low-resolution (SL module) spectrum of
  the central (3.7\arcsec\,$\times$\,3.7\arcsec) region of NGC~5135, taken
  from the {\it Spitzer} archive. The (arbitrary scaled) transmission curves
  of the IRAC/8\,$\micron$ (red, dashed line), T-ReCS/Si-2 (yellow,
  dotted-dashed line) and T-ReCS/$N$-band (blue line)
  band-pass filters are overplotted. This figure shows that although
  the filter profiles are quite different, the flux densities obtained
  with the IRAC/8\,$\micron$ (red circle) and the T-ReCS/Si-2 and N-band
  filters (yellow and (inverted) blue triangles, respectively) are not.
  [\textit{See the
      electronic edition of the Journal for a color version of this
      figure.}]}\label{f:bands} 
\vspace{.25cm}
\end{figure}

Tables~\ref{t:nucphot} and \ref{t:hiiphot} give a summary of the
aperture photometry for the LIRG nuclei, as well as for the \HII\
regions, respectively, for both the smallest and largest apertures.
Meaningful comparisons can be made with 
$Spitzer$-based works since the conversion factor between the T-ReCS
8.7\,$\micron$ and the $Spitzer$/IRAC 8\,$\micron$-band flux densities 
is close to unity for these kinds of galaxies
($\pm\,$10\%; see Fig.~\ref{f:bands} for an example). For this reason
from now on we will use the term ``8\,$\micron$''
when referring to both T-ReCS and IRAC $8\,\mu$m data.
Finally, the 8.7$\,\micron$ flux densities were converted to 
monochromatic fluxes.
Taking into account all sources of uncertainty (calibration, photometry,
and filter and color corrections, background estimation), 
the MIR photometry of the $3\sigma$ measurements have an
accuracy of $\simeq\,25$\% ($\sim\,0.1\,$dex) and
$\simeq\,40\%$ ($\sim\,0.15\,$dex)
for the $r=75\,$pc and $r=300\,$pc apertures, respectively.

\begin{figure*}
\epsscale{1.15}
\plotone{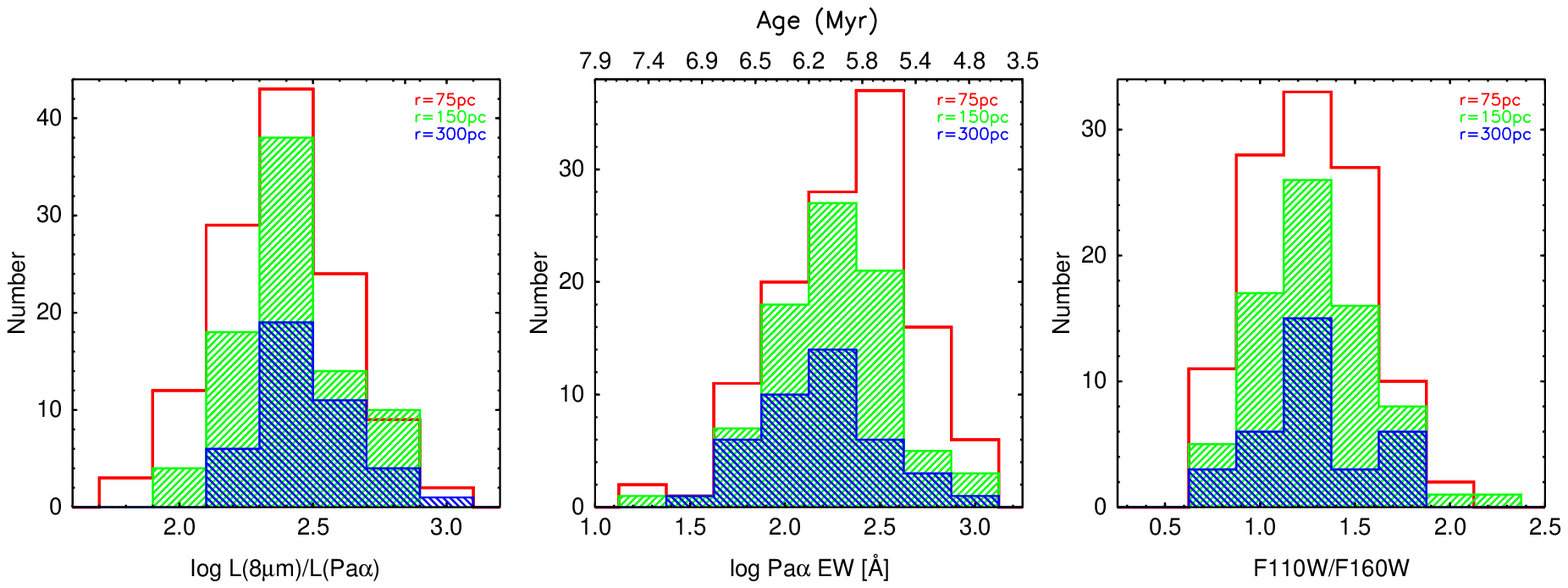}
\vspace{.25cm}
\caption{\footnotesize Distribution of the observed $L_{\rm 8\,\mu m}$/$L_{\rm
    Pa\alpha}$ ratios (left), \Paalpha\ EWs (center) and F110W/F160W continuum
  ratios (right) measured  for the \HII\ regions in our sample of LIRGs. The
    empty histogram are measurements through the $r=75\,$pc aperture, whereas 
the single and double hatched histograms are through the $r=150\,$pc 
and $r=300\,$pc apertures, respectively.  [\textit{See the electronic edition of the
      Journal for a color version of this figure.}]}\label{f:AgeExthisto} 
\end{figure*}

\subsection[]{Aperture Effects}\label{ss:apert}
Since, as discussed in \S1, in nearby galaxies the 8$\,\micron$
emission appears more extended than the \Paalpha\ (or H$\alpha$) emission 
it is important to explore the aperture effects on the measurements.
Fig.~\ref{f:AgeExthisto} (left) 
shows the distribution of the observed (not corrected
for extinction) $L_{\rm 8\,\mu m}$/$L_{\rm Pa\alpha}$ ratios 
for the three physical apertures. Although the peaks of the distributions are
similar for all of them, there is a clear tendency for the width of
the distributions to become broader for 
decreasing the aperture size. Moreover, 
the low end of the $L_{\rm 8\,\mu m}$/$L_{\rm
  Pa\alpha}$  distribution for the 75\,pc radius aperture disappears for
the large aperture distribution. 

The behavior of the \Paalpha \ EWs (Fig.~\ref{f:AgeExthisto},
middle) is different, with the peaks of the distributions moving
 toward smaller EW for increasing physical apertures. This can be
readily understood if the larger physical apertures progresively
include more continuum emission not associated with the 
\HII\ regions and/or average together young and {\it old} \HII\ regions. This may also indicate that 
the typical sizes of the \HII\ regions of LIRGs are smaller than the largest physical
aperture used here (see Alonso-Herrero et al. 2002, and e.g., Fig.~3).

Finally, the F110W/F160W distributions 
appear to be similar for the three physical apertures. 
That is, the  selected \HII\ regions in our sample of LIRGS do not 
appear to have systematically
redder NIR colors (that is, they do not appear to be more extincted; see below)
than other regions in the central parts of the galaxies.

\subsection[]{Comparison With Models}\label{ss:models}

The \textit{HST}/NICMOS imaging data can be compared with evolutionary
synthesis models of stellar populations to derive some
physical properties of the selected \HII\ regions and nuclei of the
galaxies. We ran Starburst99 (SB99, version 5.2; \citealt{Leitherer99}) for
an instantaneous burst of 10$^6$\,\Msun, Geneva tracks with high mass-loss
rate (see \citealt{Vazquez05}), Kroupa initial mass function (exponents
of 1.3 and 2.3 for 0.1\,\Msun\,$<$\,M\,$<$\,0.5\,\Msun\ and
0.5\,\Msun\,$<$\,M\,$<$\,120\,\Msun\
mass intervals, respectively), and solar metallicity. The model outputs
were obtained with a 0.5\,Myr step for starburst ages ranging from
0.5 to 10\,Myr (older stellar populations are unlikely to be detected
in \Paalpha\ emission; see \citealt{AAH02} for a detailed discussion).

\subsubsection[]{Ages}

The EW of hydrogen recombination emission lines are a useful age indicator
of young stellar populations. We compared the observed Pa$\alpha$ EW with the
model predictions to estimate the ages of the LIRG \HII\ regions.
We note that the derived ages are upper limits
to the real ages since the presence of an older underlying stellar
population would increase the
NIR continuum, thus ageing the region.


Fig.~\ref{f:AgeExthisto} (center) shows the distribution of the inferred ages
as a function of the physical aperture.
For the smallest aperture
($r=75\,$pc), the majority of the LIRG \HII\ regions 
have ages ranging from $\sim$\,5.4 and
6.8\,Myr, but star-forming regions as young as $\sim$\,4\,Myr
can be found. This is in agreement
with findings for other LIRGs (\citealt{AAH02};
\citealt{Wilson06}; \citealt{DS07a}).
When using a larger aperture, the ages inferred for the star-forming
regions tend to be higher and only a few of them appear to be 
younger than $\sim\,$5.5\,Myr (see discussion in \S\ref{ss:apert}).


\subsubsection{Extinctions}

If the age of the stellar populations is known (see above),
one can compare the observed NIR continuum colors with the model
predictions to get an estimate of the obscuration to the stars. 
To do so, the SB99 spectra were convolved (for all the starburst
ages) with the corresponding NIR filter band-passes and telescope
system throughput (F110W, F160W or F222M, depending on each galaxy)
to obtain the flux densities predicted by the models at those
wavelengths. Then, the \textit{synthetic} colors (F110W/F160W
or F160W/F222M) were interpolated to the ages of each \HII\ region
(known beforehand) and compared with the observed values.
The obscuration was calculated by using the \cite{Calzetti00}
extinction law with a foreground dust screen configuration.
The derived extinctions were then used to correct the \Paalpha\
luminosities for obscuration as well as the MIR emission (using Rieke
\& Lebofsky 1985 extinction law, since the Calzetti et al.  law extends
only up to the $K$ band).
Note that because of the ages obtained from \Paalpha\ EWs are upper
limits to the real ages, the extinctions may be lower limits to the
real ones.

Regarding the smallest aperture, the majority of the \HII\ regions
show extinctions ranging from A$_V$\,$\sim$\,4 to 8\,mag
(equivalent of Pa$\alpha$ extinctions of A$_{1.875\mu m}$\,$\sim$\,0.58--1.15\,mag, \citealt{Calzetti00}),
although there are some regions with even higher values. The median value of
the extinction for the \HII\ regions of our sample of 
LIRGs is $A_V=6\,$mag. 
The obscuration to the stars inferred in this work 
are slightly higher than those estimated by AAH06a, but entirely attributable to the use
of different models (AAH06a used the 
\citealt{Rieke93} models). 

The extinction of the LIRG \HII\ regions 
are consistent with those derived for other 
LIRGs  (\citealt{DS07a}, \citealt{Pollack07}). The extinctions of 
the \HII\ knots of the SINGS galaxies (e.g., \citealt{Cal05};
Cal07) are in agreement with the lower limit found for our \HII\ regions
($A_V \sim$\,3--4\,mag), as expected if the obscuration increases with the
SF rate, i.e., with \Paalpha\ luminosity (\citealt{Choi06} from
optical/MIR; Cal07 from NIR data). This is discussed in more detail in \S5.2.3.


The derived extinctions of the nuclei are generally higher 
than those of the \HII\
regions in the majority of the LIRGs, except for NGC~1614 (which shows a
very ``blue'' nucleus), IC~4687, IC~4734 and NGC~7130, all classified
as \HII-type or LINER. The most extinguished region in our sample is
the southern nucleus of NGC~3256 with an estimated extinction of
$A_V \sim$\,11.7\,mag in agreement with findings of other authors
(\citealt{Kotilainen96}; \citealt{Lira02}; AAH06a).

\section[]{Sub-arcsecond Morphology: Overall Characteristics}\label{s:morph}

Fig.~\ref{f:panel1} shows that the overall morphologies of the 
MIR emission and the \Paalpha\ emission line of LIRGs are  similar.
The MIR and \Paalpha\
emissions (which trace the ionizing stellar populations) are
rather concentrated in \HII\ regions and knots with sizes
of a few hundred pc, with half of the LIRG sample showing
\Paalpha\ only in the central $1-2\,$kpc, 
and the other half with \Paalpha\ emission extending
over at least $3-7\,$kpc (AAH06a for more details). 
In contrast, the morphology of the 
NIR continuum emission (which traces in general more evolved stellar
populations, see Alonso-Herrero et al. 2002) differs substantially from
that of \Paalpha\ and the MIR, showing the presence
of unresolved star clusters and diffuse emission over the whole NIC2 FOV. 

A detailed inspection of the MIR and \Paalpha\ images 
shows, however, that the spatial coincidence between them is not perfect.
There exist examples of this in every galaxy
(see Fig.~\ref{f:panel1}, center and right panels),
e.g.: the north-west part of NGC~2369, the
southern nucleus of NGC~3256, an \HII\ region located to the south-east
of the nucleus of NGC~5135, or the nuclear and some outer regions of
IC~4687. 

Fig.~\ref{f:panel2} already gives some hints about the possible 
causes giving rise to the observed differences between the MIR 
and the \Paalpha\ emission. The distribution of the cold dust, as  
traced by the F110W/F160W (or F160W/F222M) ratio
(left panel), does not seem to be correlated with the
$L_{\rm MIR}$/$L_{\rm Pa\alpha}$ maps (right) nor with the
MIR emission itself (Fig.~\ref{f:panel1}, right). This would suggest
that the dust causing the extinction to the stars might not be
related with the dust responsible of the MIR emission.
On the contrary, 
the age, as traced by the \Paalpha\ EW (center), appears to have an
important role, with the youngest 
\HII\ regions being associated with the lowest 
$8\,\mu$m to Pa$\alpha$ ratios. 
There are many good examples of this aswell, like the nuclear regions of
NGC~1614 or NGC~3256 where youngest regions are also identified
with ``holes'' in the $L_{\rm MIR}$/$L_{\rm Pa\alpha}$ images.
The nuclear region of IC~5179 also shows this behavior.
There are some cases where this anti-correlation is not so clear.
For example, the central region of IC~4687 on the other hand is not
very old but show high values of the MIR/\Paalpha\ ratio perhaps
indicating, larger contribution of the 8.6$\,\micron$ PAH feature to the
8$\,\micron$ luminosity than in other galaxies, and/or higher extinction.

\section[]{The 8$\,\micron$ \lowercase{vs.} Pa$\alpha$ Relationship}\label{s:MIRandPaa}

In this section we explore in detail the LIRG 
$8\,\mu$m vs. Pa$\alpha$ relation on
different spatial scales, compare it with the work of Cal07 
for the SINGS \HII\ knots, 
and discuss the mechanisms giving rise to the observed scatter.

\subsection[]{The $L_{\rm 8\,\mu m}$ vs. $L_{\rm Pa\alpha}$ Relation on 
Different Spatial Scales}\label{ss:MIRvsPaa}

\begin{figure}[!h]
\epsscale{1.15}
\vspace{.5cm}
\plotone{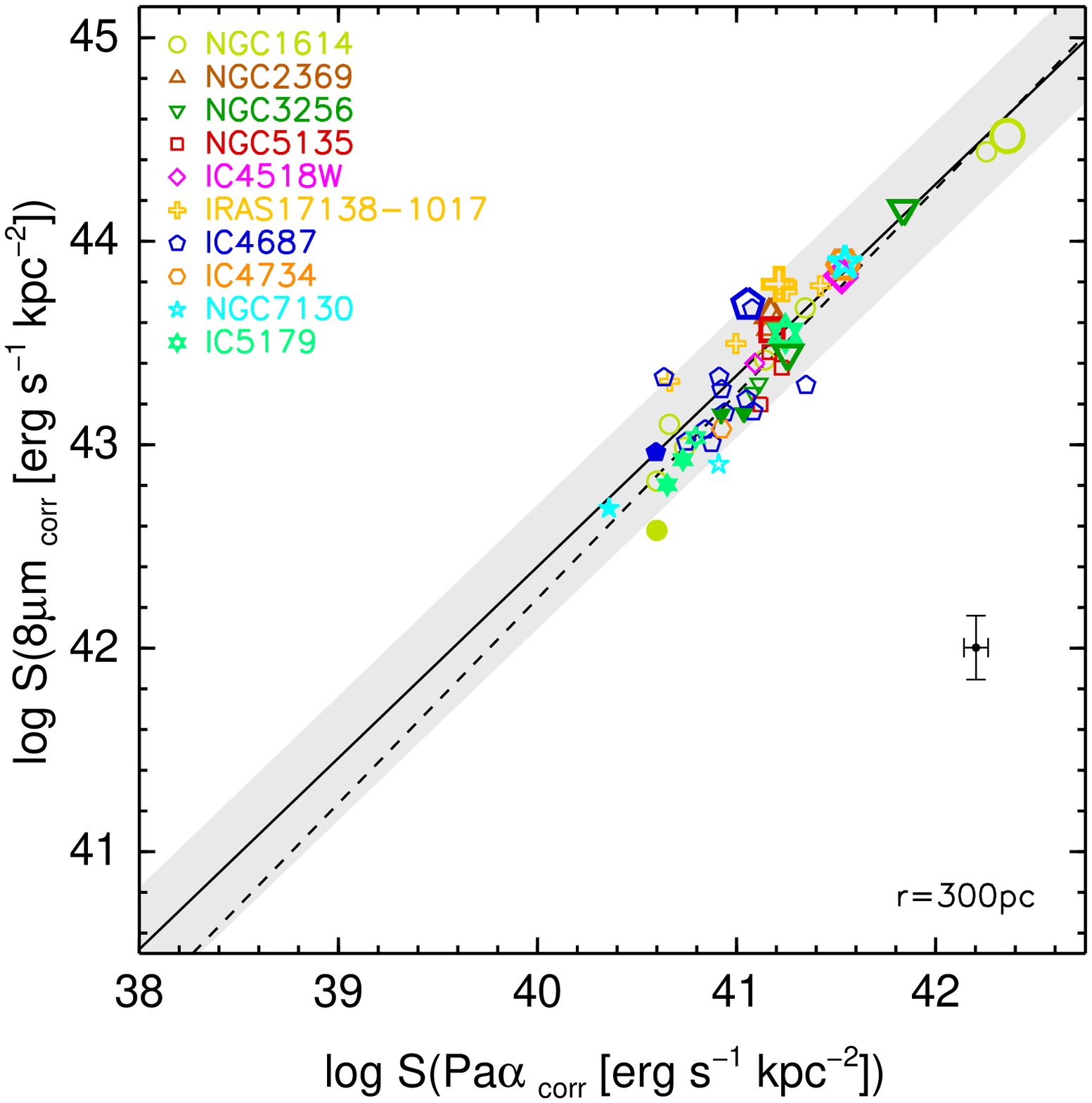}
\caption{\footnotesize The 8$\,\micron$ vs. \Paalpha\ LSD 
relation 
 (corrected for extinction) of the LIRG \HII\ regions (small open
and filled symbols are for $> 3\,\sigma$ and $2-3\,\sigma$  measurements,
respectively)
  and nuclei (big open symbols). The photometry was measured with an aperture
  of radius 300\,pc. The typical uncertainty for the $> 3\,\sigma$ measurements
  is marked with error bars.  
The solid line is
  the extrapolation of the Cal07 fit to the SINGS 
high-metallicity \HII\ knots, whereas the dashed line is our least-square
  fit to the LIRG  \HII\ regions (that is, the nuclei are excluded). 
The shaded region shows
 the $\pm 0.3$dex dispersion to the  
8\,$\micron$ vs. \Paalpha\ LSDs fit found by Cal07 for the 
  high-metallicity \HII\ knots in the SIGNS galaxy sample. Note that 
SINGS knots have LSD below $S(\Paalpha_{\rm
  corr})= 10^{40.5}\,$erg\,s$^{-1}$\,kpc$^{-2}$.
[\textit{See the electronic edition of the Journal for a color
      version of this figure.}]}\label{f:MIRvsPaa_LSD} 
\end{figure}

\begin{figure}[!h]
\epsscale{1.15}
\vspace{.5cm}
\plotone{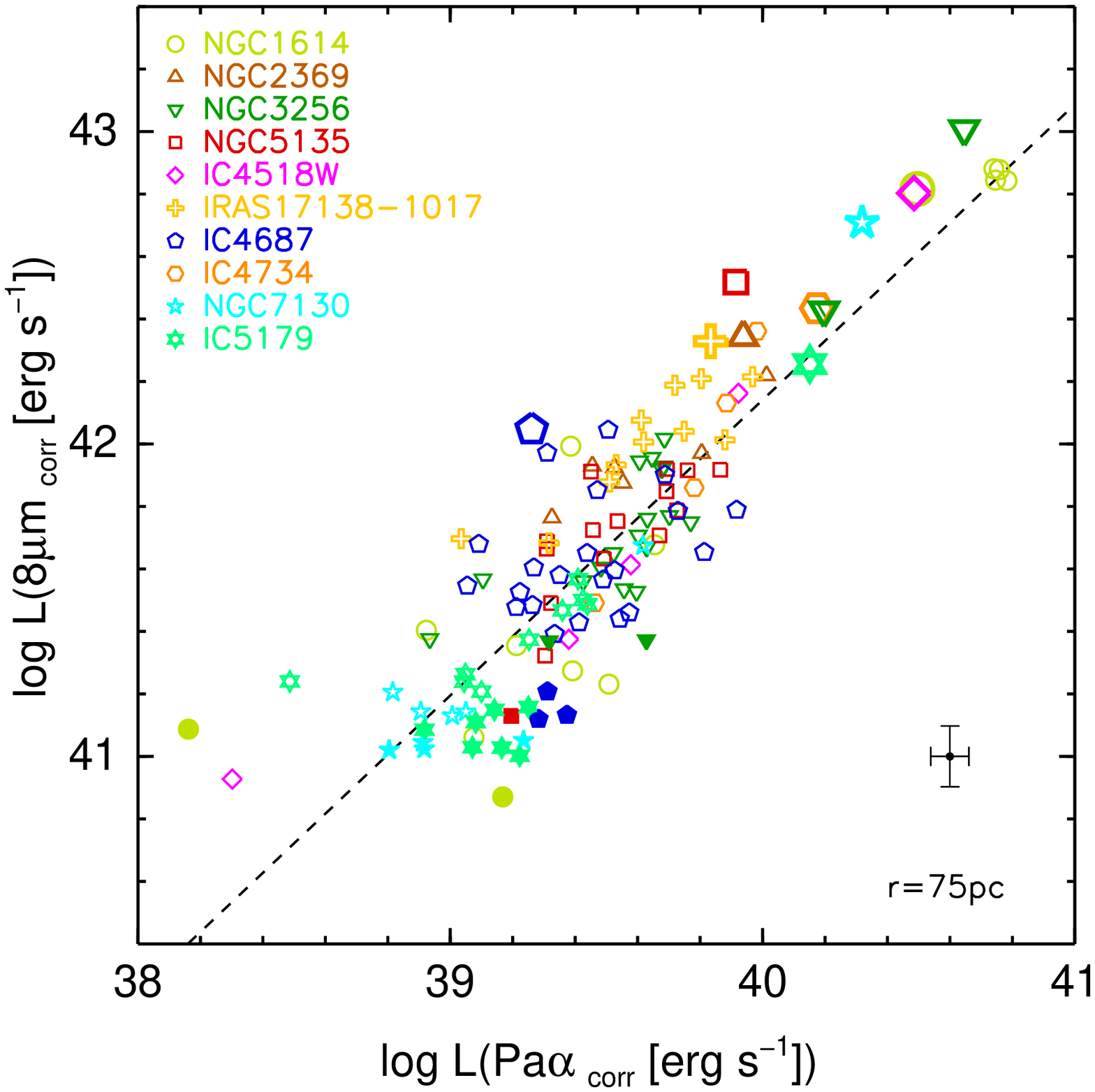}
\caption{\footnotesize The 8$\,\micron$ vs. \Paalpha\ luminosity relation
 (corrected
  for extinction) of the LIRG \HII\ regions and nuclei measured with the 
$r=75\,$pc aperture. Symbols are as in
  Fig.~\ref{f:MIRvsPaa_LSD}. The dashed 
line is our least-square fit to the
  LIRG \HII\ regions (including the $2-3\,\sigma$ measurements). [\textit{See
the electronic edition of the Journal for a color version of this
      figure.}]}\label{f:MIRvsPaa} 
\end{figure}

Fig.~\ref{f:MIRvsPaa_LSD} shows that there is a good correlation between the 
$8\,\mu$m and Pa$\alpha$ emissions for our sample of LIRG \HII\ regions when
measured through the large aperture ($r=300\,$pc). 
For an easier comparison with the work of Cal07 this figure is shown in 
units of luminosity surface density (LSD). It is clear from this figure that
the LIRG \HII\ regions (which have 12\,+\,log(O/H)\,$\sim$\,8.8;
see Table~\ref{t:char}) extend  the correlation 
found by Cal07 for the SINGS high-metallicity
(8.5\,$\lesssim$\,12\,+\,log(O/H)\,$\lesssim$\,8.9) \HII\ regions by about
two orders of magnitude above 
$S({\rm Pa}\alpha_{\rm corr}) = 10^{40.5}\,$[erg\,s$^{-1}$\,kpc$^{-2}$].
A least-square fit to our data indicates that the 
$L_{\rm 8\,\mu m}$ vs. $L_{\rm Pa\alpha}$ relationship for the 
$r=300\,$pc aperture is consistent
with a unity slope ($1.01 \pm 0.08$). The fitted slope is also
consistent with that inferred for the SINGS
high-metallicity knots ($0.94\pm0.02$, see Cal07) and that for 
the integrated  properties of galaxies ($0.92\pm 0.05$, see Wu et al. 2005). 
The LIRG \HII\ regions, however, 
seem to present a lower scatter ($\pm 0.1$dex) around the fit than the
SINGS \HII\ regions ($\pm 0.3$dex, Cal07). We attribute the smaller scatter of
our relation to the fact that
we are using the same physical sizes, whereas the Cal07 aperture photometry 
is dictated by the angular resolution of their MIPS $24\,\mu$m images. 
Their fixed 13\arcsec-diameter aperture implies physical sizes of between
$\sim 220\,$pc and
$1.3\,$kpc for their high-metallicity sample although, as explained
in \S3.1, about half of their \HII\ regions are measured through a 
$\sim600\pm100\,$pc-diameter aperture.

The 8\,$\micron$ vs. \Paalpha\ relation  holds when using 
our smallest physical aperture (see Fig.~\ref{f:MIRvsPaa}, $r=75\,$pc), 
although with a slightly smaller slope ($0.95\pm0.03$) than, but compatible
with, 
that derived for the large aperture. The main difference with the 
relation for the large aperture is the higher dispersion
around the fit ($\pm 0.2$dex), not surprising 
since the distribution of the observed MIR to Pa$\alpha$ ratios is much
broader for the small aperture (Fig.~5, \S\ref{ss:apert}). We note that this
scatter is real, as the uncertainties associated with the background estimation
are smaller for the $r=75\,$pc aperture than for the $r=300\,$pc aperture. 
The slight trend
for the slope of the $L_{\rm 8\,\mu m}$ vs. $L_{\rm Pa\alpha}$ relation
to increase with the size of the aperture was
already seen by AAH06b when comparing the relation for individual  \HII\
regions and integrated emission.

The nuclei of LIRGs classified as an AGN (NGC~5135, IC~4518W, NGC~7130)
departure from the correlations (see also AAH06b), 
and are located above the fitted relations.
All of them show an excess of MIR emission with respect to their
\Paalpha\ luminosity. This could be attributed to an additional
continuum emission due to hot dust heated by the AGN. This component
would compensate the absence of PAH emission, whose carriers are
know to be destroyed in the vicinity of a hard radiation fields,
like in AGNs (\cite{Roche85}; \cite{Voit92}; \cite{Siebenmorgen04}).
Another explanation for this excess could be
attributed to the uncertainty of the extinction corrections
as they are based on stellar colors that may not be representative
of those of an AGN. The regions identified 
as the nuclei of IRAS~17138-1017 and IC~4687
(both classified as \HII-type) also deviate from the mean trend. In both cases
the most likely explanation is that the extinction has been underestimated.

\subsection[]{What Causes the Scatter?}\label{ss:scatter}

Some caveats should be taken into account when interpreting
Figs.~\ref{f:MIRvsPaa_LSD} and \ref{f:MIRvsPaa}.
Assuming that the stellar light component is negligible at 
$\lambda > 5\,\mu$m (for high-metallicities galaxies Cal07 found that the
stellar contribution to the $8\,\mu$m emission is small),
there are two main mechanisms contributing to the $8\,\mu$m emission. 
First, there is thermal continuum from hot dust
heated by young stars (or an AGN). Obviously hot dust emission heated by an AGN is 
ruled out in \HII\ regions and \HII-like nuclei. 
The second contribution is from PAHs which could vary from galaxy to
galaxy. For instance,  \cite{Smith07} found variations in the contribution of the 
8.6\,$\micron$ PAH feature to the total PAH luminosity ($\sim 5-10$\%) and the 
total IR luminosity ($\sim$\,0.5--2\%) among the SINGS star-forming galaxies.

Cal07 modeled the MIR (both $8\,\mu$m and $24\,\mu$m) vs. Pa$\alpha$ empirical
relations for the SINGS high-metallicity 
\HII\ knots with an instantaneous burst of 4\,Myr, or equivalently 
constant star formation of
100\,Myr, and an empirical relation between extinction and the 
Pa$\alpha$ LSD. 
 The deviations from the model prediction for the high-metallicity 
\HII\ knots are explained in terms of secondary effects such as the age of the stellar
population and/or fixed extinction effects. 
Since the metallicity is almost constant among the LIRGs of
the sample, we rule it out as the main cause of the scatter in the 
$L_{\rm 8\,\mu m}$ vs. $L_{\rm Pa\alpha}$ relation.
In the following subsections, we investigate the other 
effects for our sample of LIRG \HII\ regions.

\begin{figure*}
\epsscale{1.15}
\plotone{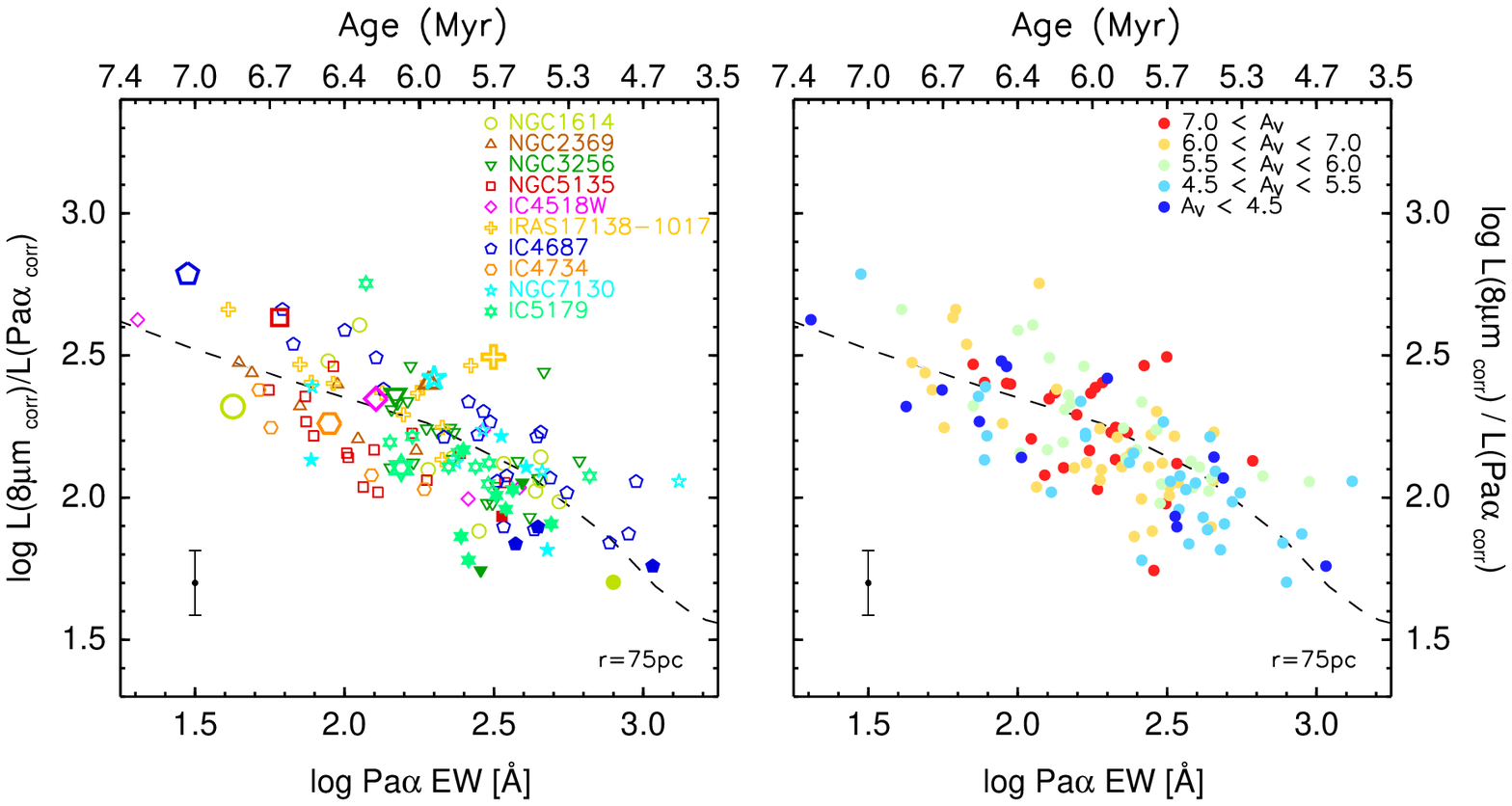}
\caption{\footnotesize Left panel: The 8$\,\micron$/\Paalpha\ luminosity ratio
  of the \HII\ regions as a function of their age (from \Paalpha\ EW
  measurements and SB99 models). Symbols are as in
  Fig.~\ref{f:MIRvsPaa_LSD}. The general evolution seen for the data-points is
  fully accounted by the model (dashed line). We assumed that 
$L_{\rm  8\mu m}/L_{\rm Pa \alpha} 
\propto L_{\rm IR}/L_{\rm Pa \alpha} \approx L_{\rm
    bol}/L_{\rm Pa\alpha}$ where
 the SB99 $L_{\rm bol}$ is scaled with a factor of $-1.59$\,dex ($\sim$\,2.6\%) to match
  the data. Right panel: Same as left panel but color coded in terms
  of the derived extinctions. [\textit{See the electronic edition of the
      Journal for a color version of this figure.}]}\label{f:MIRPaavsPaaEW} 
\end{figure*}

\subsubsection[]{The Age}

Fig.~\ref{f:MIRPaavsPaaEW} (left) explores the dependence of the 
$L_{\rm 8\,\mu m}$/$L_{\rm Pa\alpha}$ ratio (corrected for extinction) on the
age of the \HII\ regions identified in LIRGs. There
is a clear tendency for the youngest \HII\ regions in our sample to show 
the lowest $8\,\mu$m/Pa$\alpha$ ratios. This result is not subject to
aperture effects since the same trend is observed if the large aperture
is used, although somewhat diluted as 
the youngest \HII\ regions are averaged together with older regions and/or
continuum regions (see
Fig.~\ref{f:AgeExthisto}, center). Another possibility would be that the
extinction was systematically underestimated for the most obscured star-forming regions
(Rigby \& Rieke 2004). This does not seem to be the case, as there is 
no trend for the regions with high $L_{\rm 8\,\mu m}$/$L_{\rm Pa\alpha}$
ratios to correspond with \HII\ regions with the highest extinctions
(see Fig.~\ref{f:MIRPaavsPaaEW}, right).

We can use the starburst models described in 
\S3.3 to try and reproduce the trend seen in 
Fig.~\ref{f:MIRPaavsPaaEW}. Since 
in star forming galaxies the MIR monochromatic luminosities
are related to their IR luminosities
(e.g., \citealt{Elbaz02}, \citealt{Takeuchi05}; AAH06a),
and in LIRGs the IR luminosity accounts for
the majority of the bolometric luminosity,  
we can assume $L_{8\mu{\rm m}} \propto L_{\rm IR} \approx L_{\rm bol}$.
The hydrogen recombination line fluxes are, in turn, directly related
to the number of ionizing photons provided by SB99
($L_{{\rm Pa}\alpha} \propto N_{\rm Ly}$).
Taken into account these considerations, the dashed line in
Fig.~\ref{f:MIRPaavsPaaEW} represents the evolution of the
$L_{\rm bol}/L_{\rm Pa\alpha}$ ratio as a function of
the age of the starburst
as predicted by SB99 and scaled to our data-points
with a $L_{\rm 8\,\mu m}$/$L_{\rm bol} = 0.026$ ratio.
The scaling factor was calculated by means of a least-square
minimization method. As can be seen from this figure,
the general trend is well reproduced by this simple model without any previous
assumptions about the mechanisms producing  the 8\,$\micron$ emission.
In fact, the mod<el accounts for the observed variation (about one order
of magnitude) of the $L_{\rm 8\,\mu m}$/$L_{\rm Pa\alpha}$ ratio
as the starburst ages from $\sim$\,4 to 7.5\,Myr.

\subsubsection{PAH Contribution}

Although the general tendency seen in  
Fig.~\ref{f:MIRPaavsPaaEW} is accounted for by the age evolution
of the star-forming regions, a significant (vertical) scatter remains
($\pm 0.2\,$dex) around the model predictions for a given age. 
We propose that this scatter might be indeed
caused by the different contribution of the 8.6$\,\micron$ PAH
(or PAHs, depending on the filter used) to the integrated
IR emission of the galaxies. 
Fig.~\ref{f:MIRPaavsPaaEW} (left) shows that the \HII\ regions of a 
given galaxy seem to follow the model prediction with the age, 
but different galaxies appear to be shifted along the vertical 
direction (see, for example,
the \HII\ regions of NGC~5135 and those of IC~4687). This would suggest
 that the overall PAH emission could vary from galaxy to galaxy as a
whole (as seen by  \citealt{Smith07}). 
There is also the possibility 
that the PAH spectra vary from \HII\ region to \HII\ region 
within a galaxy. However, \cite{Peeters04} found that 
the variation of the $6.2\,\mu$m PAH feature of 
Galactic compact \HII\ regions is smaller than that observed in
other galaxies.

As  mentioned above, the dashed line in Fig.~\ref{f:MIRPaavsPaaEW}
shows the age evolution of the $L_{\rm bol}$/$L_{\rm Pa\alpha}$ ratio
predicted by the model, scaled with a factor of $-1.59$\,dex
($\simeq  0.026$). This factor, that comes naturally from the data,
can be interpreted as the mean contribution of the 8$\,\micron$
luminosity (dust continuum + 8.6$\,\micron$ PAH) to the
$L_{\rm bol}$ in our LIRGs.
For a sample of 59 star-forming galaxies of the SIGNS sample,
\citet{Smith07} showed that the contribution of the 8.6$\,\micron$ PAH
to the total (integrated) IR luminosity of these galaxies is in the range
of $\sim\,$0.5 to 2\%. Since our 8$\,\micron$ luminosities account
not only for the 8.6$\,\micron$ PAH but also for the continuum emission,
the 2.6\% factor inferred above for our LIRGs is in agreement with
their results. Moreover, the dispersion range of the 8.6$\,\micron$ PAH
intensity with respect to the $L_{\rm IR}$ among the SINGS galaxies,
$\pm 0.3\,$dex, is also in agreement with the (vertical) scatter
seen in our data, suggesting the variation of the overall PAH emission
field among the LIRGs as to be the main contributor to this scatter.

\subsubsection{Extinction vs. Ionizing Flux}\label{ss:extion}

\begin{figure}
\epsscale{1.15}
\plotone{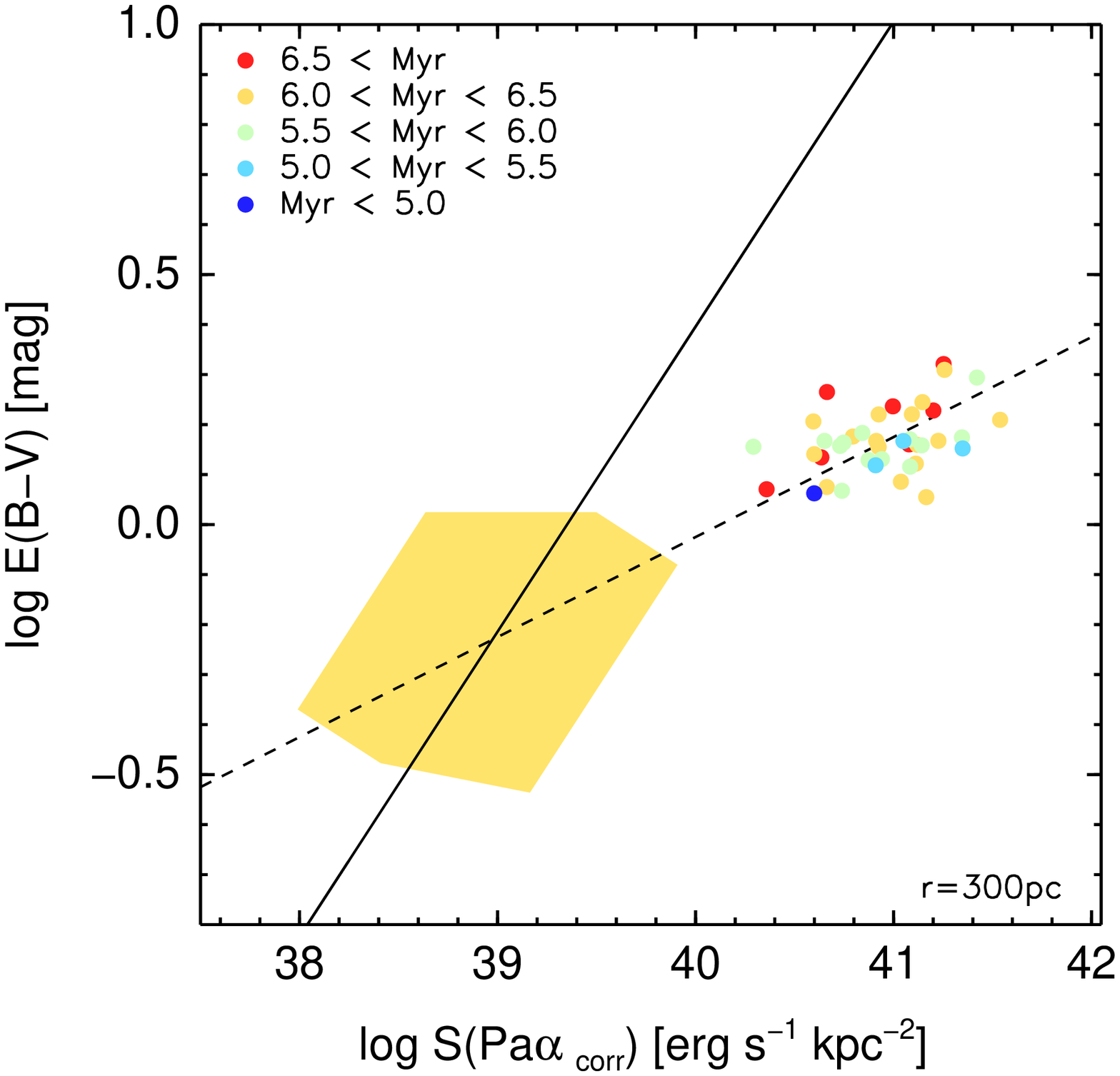}
\caption{\footnotesize Color excess $E(B-V)$ of the LIRG \HII\ regions as a function
  of their extinction-corrected \Paalpha\ LSD (for the $r=300\,$pc data). The figure is color coded
  to verify that there is no trend with the age of the regions. The solid line
  is the fit of Cal07 to their high-metallicity \HII\ knots as given in their
  equation~A2. The shaded region is the \Paalpha\ LSD and extinction
  range of their data. To guide the eye we have drawn the dashed line
passing through both LIRG \HII\ regions and the SINGS high-metallicity \HII\ knots (from their Fig.~12). [\textit{See
      the electronic edition of the Journal for a color version of this
      figure.}]}\label{f:ExtvsPaa} 
\end{figure}

Although we find that 
the scatter of the $L_{\rm 8\,\mu m}$/$L_{\rm Pa\alpha}$ ratio is not related
with a residual
extinction effect (Fig.~\ref{f:MIRPaavsPaaEW}, right),
Cal07 found that SINGS \HII\ knots with increasing Pa$\alpha$ luminosities tend
to show higher extinctions. Fig.~\ref{f:ExtvsPaa}, plotted in LSD units so that it
can be directly compared with Fig.~12 of Cal07, shows that such tendency is
present in our sample. We find however that the LIRG \HII\ regions do not follow the
extrapolation of the best fit of Cal07, which lies well above ($\sim\,$1\,dex
in E(B-V)) the
location of our star-forming regions. 
We note that the tendency seen for the LIRG \HII\ regions appears to be
consistent with an apparent flattening of the relation for the few data-points of Cal07 with 
$\log S(\rm Pa\alpha) \gtrsim\,10^{40}\,$erg\,s$^{-1}$\,kpc$^{-1}$.

One possibility is that we are
underestimating the extinctions, as they are derived from NIR colors. 
However, even if we assumed that $A_V$(stars)\,$\simeq$\,0.44\,$A_V$(gas)
(\citealt{Calzetti00}), this would imply a correction of only --\,0.3\,dex,
which still would not put the LIRG \HII\ regions on the extrapolation of the 
Cal07 fit. 
In fact, the extrapolation of the Cal07 fit to our
\Paalpha\ LSD regime ($\gtrsim\,10^{40.5}\,$erg\,s$^{-1}$\,kpc$^{-1}$)
would imply color excesses $E(B-V) > 10$, i.e., visual extinctions 
$A_V > 40\,$mag (for a foreground dust configuration).
In contrast, we have obtained relatively modest attenuations for
our \HII\ regions in the range of $\sim\,$4--8\,mag. 
AAH06 also used \Halpha/\Paalpha\ and \Paalpha/Br$\gamma$
line ratios to calculate the extinction to the gas and found integrated
($\approx\,$2\arcsec$\,\times\,$7\arcsec) values for the sample of LIRGs
of $A_V\,\sim\,$2--14\,mag, still well below the values we would obtain from
the extrapolation of the Cal07
fit.

\section{Summary}\label{s:conclu}

In this paper we presented T-ReCS sub-arcsecond (FWHM$\,\sim\,0.4\arcsec$)  
MIR ($8.6\,\mu$m or $10.3\,\mu$m)
imaging observations of a sample of ten low-$z$ ($d<76\,$Mpc) nearly solar
metallicity ($12 + \log({\rm O/H}) \sim 8.8$) LIRGs, as well as {\it
  HST}/NICMOS continuum and Pa$\alpha$ images. The main goal was to study in 
detail the $L_{\rm 8\,\mu m}$ vs.
$L_{\rm Pa\alpha}$ relationship for \HII\ regions in LIRGs 
on scales of a few hundred parsecs (FWHM$\sim 120\,$pc for the average
distance $d=60\,$Mpc). We performed 
photometry of \HII\ regions through apertures with 
radii of $r=75\,$pc (122 \HII\ regions), $r=150\,$pc (84 \HII\ regions) and $r=300\,$pc
(49 \HII\ regions). The first aperture was chosen to take advantage of
the high angular resolution afforded by T-ReCS and NICMOS. The large
aperture  is useful to compare our 
results with those of  Cal07 for \HII\ regions in the high-metallicity  SIGNS
galaxies observed with {\it Spitzer}/IRAC at $8\,\mu$m.

We find that although the overall Pa$\alpha$ (tracing the youngest ionizing
stellar populations) morphologies of LIRGs are similar to those in the MIR,
there are some differences on the $\sim 100\,$pc scales. The morphological
differences appear to be related to the age of the young 
stellar populations, with regions of low 
$L_{\rm 8\,\mu m}/L_{\rm Pa\alpha}$ ratios showing large 
Pa$\alpha$ EW. In general we do not find a relation between  
red NIR colors, which would indicate high extinction to the stars, 
and regions of high or low $L_{\rm 8\,\mu m}/L_{\rm Pa\alpha}$ ratios.

On scales of  $r=300\,$pc the LIRG \HII\ regions extend  the SINGS 
$L_{\rm 8\,\mu m}$ vs. $L_{\rm Pa\alpha}$ 
relation by about two orders of magnitude (as already found by AAH06b)
for Pa$\alpha$ LSD above $10^{40.5}\,$erg\,s$^{-1}$\,kpc$^{-2}$.  
When studied on the small scales ($r=75\,$pc) the relation holds, although  with a  
slightly shallower slope and  a greater (real) scatter around the fit ($\pm
0.2\,$dex).  Taking into account that our sample has a nearly constant
metallicity, the scatter of this relation 
is explained in terms of the ages of 
the ionizing population and different PAH contributions. There is a
tendency for the youngest \HII\ 
regions in our sample to show low $L_{\rm 8\,\mu m}/L_{\rm Pa\alpha}$ ratios. 
Considering instantaneous star formation and assuming that
$L_{\rm 8\mu m}\,\propto\,L_{\rm IR}\,\approx\,L_{\rm bol}$, we 
naturally reproduce the observed  $L_{\rm 8\,\mu m}/L_{\rm Pa\alpha}$
ratios, which vary by a factor of ten,  with  ages  ranging 
from $\sim$\,4 to 7.5\,Myr.
 The residual dispersion around the model prediction is
likely to be caused by the different contribution from galaxy to galaxy of 
the $8.6\,\mu$m PAH feature (in our case) 
to the 8$\,\micron$ emission (and in general, to the IR
luminosity), as observationally found by Smith et
al. (2007) for the SINGS galaxies. 

Although we see a trend for the LIRG \HII\ regions with the largest Pa$\alpha$
LSD to show the highest extinctions to the stars, they
do not follow the extrapolation 
of the relation between the $E(B-V)$ color excess and the 
\Paalpha\ LSD found by Cal07, 
which would imply extinctions in excess of $A_V=40\,$mag.
In contrast, they show relatively modest attenuations.

\section*{Acknowledgements}

We thank the anonymous referee for useful comments and suggestions.
This work has been supported by the Plan Nacional del Espacio under
grant ESP2005-01480 and ESP20076-65475-C02-01. TDS acknowledges support
from the Consejo Superior de Investigaciones Cient\'{\i}ficas under
grant I3P-BPD-2004. This research has made use of the NASA/IPAC
Extragalactic Database
(NED), which is operated by the Jet Propulsion Laboratory, California
Institute of Technology, under contract with the National Aeronautics
and Space Administration, and of NASA's Astrophysics Data System (ADS)
abstract service.\\


\clearpage

\begin{deluxetable*}{lccccc}
\tabletypesize{\scriptsize}
\tablewidth{0pc}
\tablecaption{\scriptsize The Sample}
\tablehead{\colhead{Galaxy} & \colhead{z} & \colhead{Dist} & \colhead{log
    L$_{IR}$} & \colhead{Type} & \colhead{12+log(O/H)} \\ 
\colhead{name} & \colhead{(\kms)} & \colhead{(Mpc)} & \colhead{(\Lsun)} & & \\
\colhead{(1)} & \colhead{(2)} & \colhead{(3)} & \colhead{(4)} & \colhead{(5)} & \colhead{(6)}}
\startdata 
NGC 1614        & 0.01594 & 69.1 & 11.67 & \HII   & 8.6$^a$ \\ 
NGC 2369        & 0.01081 & 46.7 & 11.14 & \HII     & 8.9 \\ 
NGC 3256        & 0.00935 & 40.4 & 11.67 & \HII   & 8.8 \\ 
NGC 5135        & 0.01369 & 59.3 & 11.27 & Sy2    & 8.7 \\ 
IC 4518W        & 0.01573 & 68.2 & 11.09 & Sy2    & 8.6 \\ 
IRAS 17138-1017 & 0.01734 & 75.3 & 11.39 & \HII   & 8.9 \\ 
IC 4687         & 0.01735 & 75.3 & 11.55: & \HII  & 8.8 \\ 
IC 4734         & 0.01561 & 67.7 & 11.28 & \HII/L & 9.0 \\ 
NGC 7130        & 0.01615 & 70.1 & 11.39 & L/Sy   & 8.8 \\ 
IC 5179         & 0.01142 & 49.3 & 11.20 & \HII   & 8.9    
\enddata
\tablecomments{\scriptsize (1) Galaxy; (2) Redshift (NED); (3) Distance as obtained with the cosmology: $H_0=70\,{\rm km\,s}^{-1}{\rm Mpc}^{-1}$, $\Omega_{\rm M}=0.27$, $\Omega_\Lambda=0.73$; (4) Infrared luminosity as computed from $IRAS$ fluxes (\citealt{Sanders03}) and with the prospect given by Sanders \& Mirabel (1996) (their Table 1); (5) Nuclear activity of the galaxy; (6) Oxygen abundance taken from \cite{Relano07} and $^a$ \cite{Vacca92}}\label{t:char}
\end{deluxetable*}

\begin{deluxetable*}{lcccc}
\tabletypesize{\scriptsize}
\tablewidth{0pc}
\tablecaption{\scriptsize Observation Details}
\tablehead{\colhead{Galaxy} & \colhead{Filter} & \colhead{t$_{int}$} & \colhead{Date} & \colhead{Seeing} \\
\colhead{name} & & \colhead{(s)} & & \colhead{(\arcsec)} \\
\colhead{(1)} & \colhead{(2)} & \colhead{(3)} & \colhead{(4)} & \colhead{(5)}}
\startdata 
NGC 1614        & Si-2 & 1680     & 09/16-30/2006      & 0\farcs38 \\
NGC 2369        & Si-2 & 840      & 09/30/2006         & 0\farcs38 \\
NGC 3256        & Si-2 & 300      & 03/04/2006         & 0\farcs30 \\
NGC 5135        & $N$  & 600      & 03/04/2006         & 0\farcs31 \\
IC 4518W        & $N$  & 1200     & 04/10-18/2006      & 0\farcs33 \\
IRAS 17138-1017 & Si-2 & 840      & 09/09/2006         & 0\farcs41 \\
IC 4687         & Si-2 & 840      & 09/09/2006         & 0\farcs31 \\
IC 4734         & Si-2 & 840      & 09/09/2006         & 0\farcs31 \\
NGC 7130        & $N$  & 600      & 09/18/2005         & 0\farcs32 \\
IC 5179         & Si-2 & 840      & 09/28/2006         & 0\farcs33 
\enddata
\tablecomments{\scriptsize (1) Galaxy; (2) Filter with which each galaxy was
  observed; 
(3) On-source integration time; (4) Date(s) of the observations; (5) Seeing
  (FWHM of the reference 
standard star)}\label{t:obs}
\end{deluxetable*}

\begin{deluxetable}{lccccc}
\tabletypesize{\scriptsize}
\tablewidth{0pc}
\tablecaption{\scriptsize Aperture photometry of the nuclei}
\tablehead{\colhead{Galaxy} & & \multicolumn{2}{c}{$r=75\,$pc} & \multicolumn{2}{c}{$r=300\,$pc} \\
\colhead{name} & \colhead{FWHM} & \colhead{MIR} & \colhead{\Paalpha} & \colhead{MIR} & \colhead{\Paalpha} \\
 & \colhead{(\arcsec)} & \colhead{(mJy)} & \colhead{(mJy)} & \colhead{(mJy)} & \colhead{(mJy)} \\
\colhead{(1)} & \colhead{(2)} & \colhead{(3)} & \colhead{(4)} & \colhead{(3)} & \colhead{(4)}}
\startdata
\multicolumn{6}{c}{$N$ ($10.3\,\mu$m) filter} \\
\hline
NGC 5135        & 0\farcs36 &  31 & 0.55 &  95 &  3.2  \\
IC 4518W        & 0\farcs39 &  38 & 0.96 & 118 &  3.9  \\
NGC 7130        & 0\farcs45 &  38 & 1.27 & 148 &  5.5  \\
\hline
\multicolumn{6}{c}{Si-2 ($8.7\,\mu$m) filter} \\
\hline
NGC 1614        & 0\farcs63 &  28 & 2.19 & 346 & 34    \\
NGC 2369        & 0\farcs60 &  16 & 0.80 &  95 &  4.1  \\
NGC 3256        & 0\farcs47 & 109 & 6.57 & 427 & 27    \\
IRAS 17138-1017 & 0\farcs83 &   6 & 0.20 &  47 &  1.4  \\
IC 4687         & \ldots    &   4 & 0.09 &  43 &  1.4  \\
IC 4734         & 0\farcs90 &  10 & 0.70 &  80 &  4.7  \\
IC 5179         & 0\farcs87 &  13 & 1.30 &  70 &  4.7   
\enddata
\tablecomments{\scriptsize The flux densities presented in this table have not been corrected for extinction and are subject to the uncertainties explained in \S~\ref{s:obs}. MIR $N$-band flux densities are shown without the color correction applied (see \S~\ref{ss:photcal}). (1) Galaxy; (2) FWHM of the region selected as the nucleus of the galaxy; the nucleus of IC~4687 is very diffuse so no FWHM could be measured; (3) and (4) Median MIR and \Paalpha\ flux densities of the nuclei. The nuclei of NGC~5135 and IC~4518W appear as almost point sources in the MIR images (compare with Table~\ref{t:obs}). An aperture correction factor of 2 and 2.5, respectively, should be applied to the values of the smallest aperture if the flux densities of the point sources want to be obtained; Values of (3)--(4) are given for the smallest ($r=75\,$pc) and largest apertures ($r=300\,$pc).}\label{t:nucphot}
\end{deluxetable}

\begin{deluxetable}{lcccccc}
\tabletypesize{\scriptsize}
\tablewidth{0pc}
\tablecaption{\scriptsize Aperture photometry of the \HII\ regions}
\tablehead{\colhead{Galaxy} & \multicolumn{3}{c}{$r=75\,$pc} & \multicolumn{3}{c}{$r=300\,$pc} \\
\colhead{name} & \colhead{\#} & \colhead{MIR} & \colhead{\Paalpha} & \colhead{\#} & \colhead{MIR} & \colhead{\Paalpha} \\
 & & \colhead{(mJy)} & \colhead{(mJy)} & & \colhead{(mJy)} & \colhead{(mJy)} \\
\colhead{(1)} & \colhead{(2)} & \colhead{(3)} & \colhead{(4)} & \colhead{(2)} & \colhead{(3)} & \colhead{(4)}}
\startdata
\multicolumn{7}{c}{$N$ ($10.3\,\mu$m) filter} \\
\hline
NGC 5135        & 15 (1) &  3.7  & 0.29 &  3     & 42 & 3.2  \\
IC 4518W        &  4     &  1.1  & 0.11 &  1     & 32 & 1.6  \\
NGC 7130        &  9 (4) &  0.6  & 0.05 &  2 (1) & 10 & 1.2  \\
\hline
\multicolumn{7}{c}{Si-2 ($8.7\,\mu$m) filter} \\
\hline
NGC 1614        & 13 (2) &  3.6  & 0.21 &  7 (1) & 14 & 0.9  \\
NGC 2369        &  6     &  6.8  & 0.49 &  2     & 84 & 4.0  \\
NGC 3256        & 20 (2) &  6.1  & 0.64 &  5 (2) & 58 & 5.3  \\
IRAS 17138-1017 & 11     &  2.8  & 0.13 &  4     & 34 & 1.3  \\
IC 4687         & 24 (3) &  1.3  & 0.08 & 12 (1) & 16 & 1.0  \\
IC 4734         &  4     &  4.8  & 0.30 &  2     & 44 & 2.9  \\
IC 5179         & 16 (7) &  1.6  & 0.16 &  3 (2) & 20 & 1.6  %
\enddata
\tablecomments{\scriptsize The flux densities presented in this table have not been corrected for extinction and are subject to the uncertainties explained in \S~\ref{s:obs}. (1) Galaxy; (2) Final number of \HII\ regions selected in each LIRG. In parentheses are given the number of $2-3\sigma$ detections; (3) and (4) Median MIR and \Paalpha\ flux densities of the \HII\ regions selected in the T-ReCS images (see text for details); Values of (2)--(4) are given for the smallest ($r=75\,$pc) and largest apertures ($r=300\,$pc).}\label{t:hiiphot}
\end{deluxetable}

\end{document}